\documentclass[pre,aps,twocolumn,amsmath,amssymb]{revtex4}

\newcommand{\bb}{\begin{equation}}
\newcommand{\ee}{\end{equation}}
\newcommand{\ba}{\begin{eqnarray}}
\newcommand{\ea}{\end{eqnarray}}
\newcommand{\rhor}{\rho({\bf r})}
\newcommand{\dd}{{\rm d}}
\newcommand{\rr}{{\mathbf r}}
\newcommand{\dr}{{\rm d}{\bf r}}
\newcommand{\kk}{{\mathbf k}}
\DeclareMathOperator{\Si}{Si}

\makeatletter
\def\@frameeq#1{%
  \framebox{$\,\displaystyle#1\hbox{\vrule height 2.4ex depth 1.4ex width 0pt}\,$}}
\newcommand\Equation[1]{$$\refstepcounter{equation}%
  \@frameeq{#1}%
  \eqno \hbox{\@eqnnum}$$\@ignoretrue\ignorespaces}
\newcommand\Displaystyle[1]{$$\@frameeq{#1}$$\@ignoretrue\ignorespaces}
\makeatother

\usepackage{graphicx}
\usepackage{dcolumn}
\usepackage{bm}
\usepackage{longtable}

\begin{document}

\title{Phase behaviour of fluids in undulated nanopores}

\author{Martin \surname{Posp\'\i\v sil}}
\affiliation{
{Department of Physical Chemistry, University of Chemical Technology Prague, Praha 6, 166 28, Czech Republic;}\\
{The Czech Academy of Sciences, Institute of Chemical Process Fundamentals,  Department of Molecular Modelling, 165 02 Prague, Czech Republic}}                
\author{Alexandr \surname{Malijevsk\'y}}
\affiliation{ {Department of Physical Chemistry, University of Chemical Technology Prague, Praha 6, 166 28, Czech Republic;} {The Czech Academy of Sciences, Institute of Chemical Process
Fundamentals,  Department of Molecular Modelling, 165 02 Prague, Czech Republic}}

\begin{abstract}
\noindent The geometry of walls forming a narrow pore may qualitatively affect the phase behaviour of the confined fluid. Specifically, the nature of
condensation in nanopores formed of sinusoidally-shaped walls (with amplitude $A$ and period $P$) is governed by the wall mean separation $L$ as
follows. For $L>L_t$,  where  $L_t$ increases with $A$, the pores exhibit standard capillary condensation similar to planar slits. In contrast, for
$L<L_t$, the condensation occurs in two steps, such that the fluid first condenses locally via bridging transition connecting adjacent crests of the
walls, before it condenses globally. For the marginal value of $L=L_t$, all the three phases (gas-like, bridge and liquid-like) may coexist. We show
that the locations of the phase transitions can be described using geometric arguments leading to modified Kelvin equations. However, for completely
wet walls, to which we focus on, the phase boundaries are shifted significantly due to the presence of wetting layers. In order to take this into
account, mesoscopic corrections to the macroscopic theory are proposed. The resulting predictions are shown to be in a very good agreement with a
density functional theory even for molecularly narrow pores. The limits of stability of the bridge phase, controlled by the pore geometry, is also
discussed in some detail.
\end{abstract}

\maketitle

\section{Introduction}

It is well known that fluids which are subjects of narrow confinements exhibit quite different phase  behaviour compared to their bulk counterparts
\cite{rowlin, hend, nakanishi81, nakanishi83, gelb}. A fundamental example of this is a phenomenon of capillary condensation occurring in  planar
slits  made of two identical parallel walls a distance $L$ apart. Macroscopically, the shift in the chemical potential, relative to its saturation
value $\mu_{\rm sat}$, at which capillary condensation occurs, is given by the Kelvin equation (see, e.g. Ref.\cite{gregg})
 \bb
 \delta \mu_{\rm cc}^\parallel=\frac{2\gamma\cos\theta}{L\Delta\rho}\,,\label{kelvin}
 \ee
where $\gamma$ is the liquid-gas surface tension, $\theta$ is the contact angle characterizing wetting properties of the walls,
$\Delta\rho=\rho_l-\rho_g$ is the difference between the number densities of coexisting bulk liquid and gas. Here, the Laplace pressure difference
$\delta p$ across the curved interface separating the gas and liquid phases has been approximated by $\delta p\approx\delta\mu\Delta\rho$, accurate
for small undersaturation \cite{evans85}. Microscopic studies of capillary condensation based on density functional theory (DFT) \cite{evans84,
evans85, evans85b, evans86, evans86b, evans87, evans90} and computer simulation \cite{binder05, binder08} have shown that the Kelvin equation is
surprisingly accurate even for microscopically narrow pores. This is particularly so for walls that are partially wet ($\theta>0$) where the Kelvin
equation remains quantitatively accurate even for slits which are only about ten molecular diameters wide \cite{tar87}. For completely wet pores
($\theta=0$), Eq.~(\ref{kelvin}), which ignores the presence of thick wetting layers adsorbed at the walls, is somewhat less accurate but its
mesoscopic extension based on Derjaguin's correction \cite{derj} provides excellent predictions for the location of capillary condensation even at
nanoscales.

Capillary condensation in planar slits can be interpreted as a simple finite-size shift of the bulk liquid-gas transition  controlled by a single
geometric parameter $L$, which also determines a shift in the critical temperature $T_c(L)$ beyond which only a single phase in the capillary is
present \cite{evans86}. On a mean-field level, the transition can be determined by constructing adsorption (initiated at a gas-like state) and
desorption (initiated at a liquid-like state) isotherms which form low- and high-density branches of the van der Waals loop and which have the same
free energies right at the chemical potential $\mu_{\rm cc}^\parallel=\mu_{\rm sat}-\delta\mu_{\rm cc}^\parallel$.

However, the situation becomes significantly more sophisticated for pores of non-planar geometry. In those more general cases the translation
symmetry may be broken not only across but also along the confining walls, which can make the phenomenon of capillary condensation much more subtle.
For example, by considering a semi-infinite slit made of capping the open slit at one end, the transition, which occurs at the same value of
$\mu_{\rm cc}^\parallel$, can become second-order due to the formation of a single meniscus which continuously unbinds from the capped end, as the
chemical potential is increased towards $\mu_{\rm cc}^\parallel$
\cite{darbellay,evans_cc,tasin,mistura,mal_groove,parry_groove,mistura13,our_groove,bruschi2, fin_groove_prl}. If such a capped capillary is not
semi-infinite but of a finite depth $D$ (measuring a distance between the capped and the open end of the capillary), asymmetric effective forces
acting on the meniscus from both capillary ends round and shift the transition by an amount scaling with $D^3$ for systems with dispersion forces
\cite{fin_groove}.

Another example of the impact of broken translation symmetry on phase behaviour in narrow slits is when the walls are no longer smooth but are
structured chemically or geometrically. If the width of such slits is considerably larger than a length characterizing the lateral structure, the
condensation scenario will differ from that for non-structured slits just quantitatively. For instance, if the walls are formed of two species with
different contact angles, then the location of capillary condensation will be macroscopically given by Eq.~(\ref{kelvin}), in which Young's contact
angle is replaced by the effective contact angle given by Cassie's law \cite{cassie, het_slit_prl}. However, for sufficiently narrow pores the walls
structure may play more significant role and can change the mechanism of the condensation, which happens in two steps, such that the fluid first
condenses only locally by forming liquid bridges across the pore \cite{het_slit_prl, chmiel, rocken96, rocken98, bock, swain, bock2, valencia,
hemming, schoen2}.


\begin{figure}[h]
\includegraphics[width=0.5\textwidth]{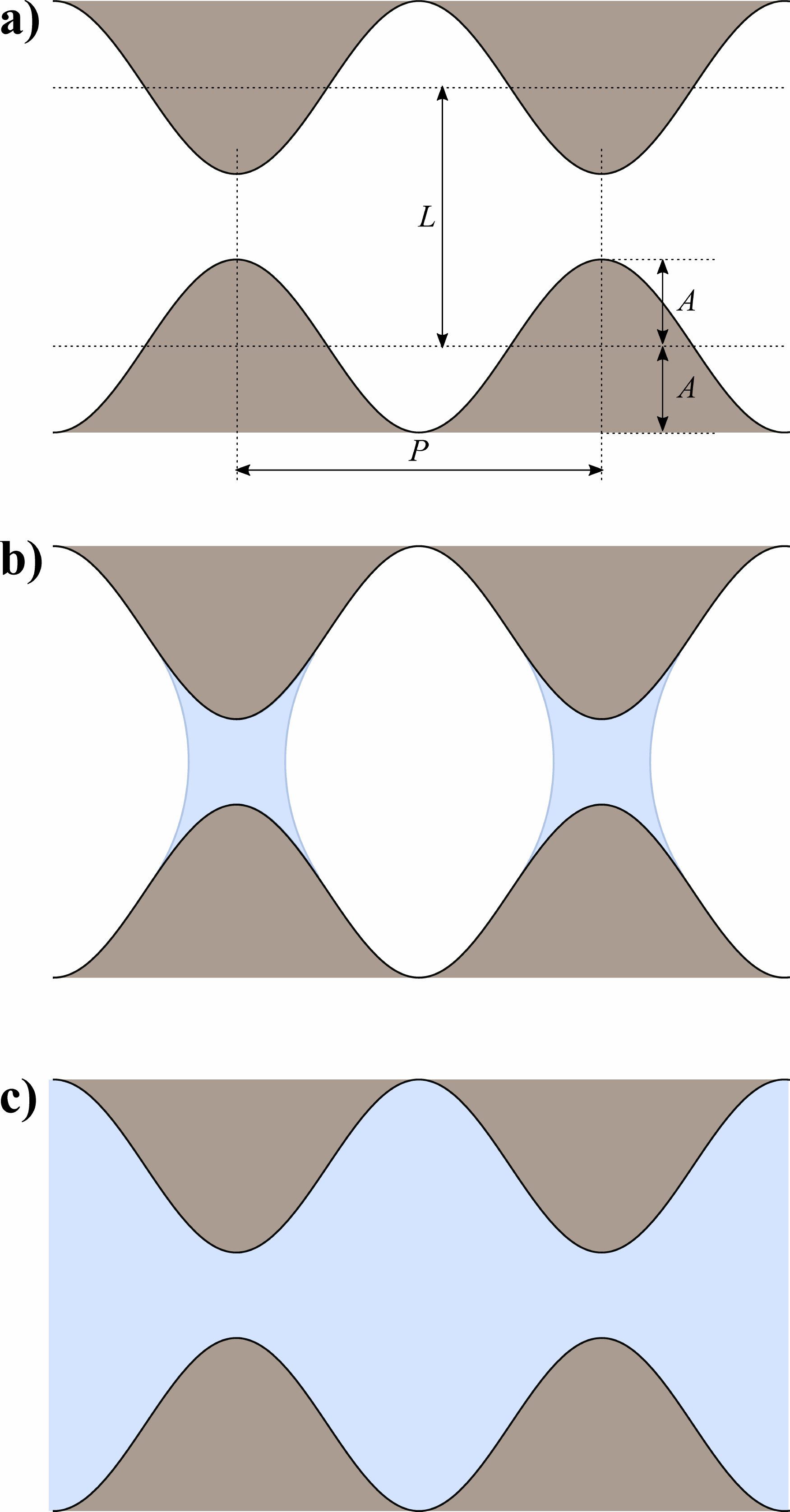}
\caption{A sketch illustrating three possible phases in a nanopore of a mean width $L$ formed  by sinusoidally-shaped walls with an amplitude $A$ and
period $P$: a) gas phase, b) bridge phase, and c) liquid phase.}\label{sketch_phases}
\end{figure}

In this paper, we study phase behaviour of fluids in pores formed of smoothly undulated and completely wet walls. The particular emphasis is put on
model pores formed by a pair of sinusoidally-shaped walls, where one of the walls is a reflection symmetry of the other. In this way, the translation
symmetry of the system is broken along two of the Cartesian axes ($x$ and $z$, say) but is maintained along the remaining one ($y$ axis). Let $P$ be
the period and $A$ the amplitude of the walls, whose mean separation is $L$.  Hence, the local width of the pore  smoothly varies (as a function of
$x$) between $L-2A$ and $L+2A$. The model, together with a macroscopic illustration of possible phases, which the confined fluid is anticipated to
adopt, is sketched in Fig.~1.

The purpose of this paper is to present a detailed analysis of the phase behaviour of (simple) fluids in such confinements. To this end, we first
formulate a purely macroscopic theory based on geometric arguments. This allows us to determine the mean separation of the walls $L_t$, which
separates two possible condensation regimes. For $L>L_t$, capillary condensation occurs in one step and macroscopically its location is given by a
trivial modification of Eq.~(\ref{kelvin}), leading to a marriage of the Kelvin equation with Wenzel's law \cite{wenzel}, such that the latter is of
the form
 \bb
 ``\cos{\theta^{*}}"=r\cos\theta\,. \label{wenzel}
 \ee
 Here $r$ is the roughness parameter of the wall and the symbol
$``\cos{\theta^{*}}"$ characterizes an enhancement of the wetting properties of the wall due to its nonplanar geometry.

In contrast, for $L<L_t$, when the condensation is a two-step process, the phase boundaries between gas-like (G) and bridge (B) phases, as well as
between bridge and liquid-like (L) phases are macroscopically determined. This requires to find how the location of the bridging films varies with
the chemical potential and we also examine the limits of metastable extensions of  B phase due to the pore geometry. Moreover, in order to capture
the effect of adsorbed wetting layers, which the purely macroscopic theory neglects, the mesoscopic corrections, incorporating the wetting properties
of the walls, are included. The resulting predictions will be shown to be in an excellent agreement with a microscopic density functional theory
(DFT), even on a molecular scale of the walls parameters.

The rest of the paper is organized as follows. In section \ref{macro} we formulate a macroscopic theory determining phase boundaries between all the
G, B, and L phases using simple geometric arguments. We start with considering a general model of a nanopore whose shape is represented by a smooth
function $\psi(x)$, before we focus specifically to sinusoidally-shaped walls. The geometric considerations are further applied to estimate the range
of stability of B phase. In section \ref{meso} we extend the macroscopic theory by including the mesoscopic corrections for walls exerting
long-range, dispersion potentials. In section \ref{dft} we formulate the microscopic DFT model, which we use to test the aforementioned predictions;
the comparison is shown and discussed in section \ref{results}. Section  \ref{summary} is the concluding part of the paper where the main results of
this work are summarized and its possible extensions are discussed.


\section{Macroscopic description of capillary condensation and bridging transition for completely wet walls} \label{macro}

\subsection{General model}

We consider a pore of a mean width $L$ formed by a pair of walls each of shape $\psi(x)$, where $x$ is a horizontal axis placed along the pore, such
as in Fig.~2. More specifically, the vertical heights of the top and bottom walls measured along the $z$-axis are $z_w(x)$ and $-z_w(x)$,
respectively, with
 \bb
 z_w(x)=\frac{L}{2}-\psi(x)\,, \label{zw_gen}
 \ee
assuming that $\psi(x)$ is a differentiable, even and periodic function of wavelength $P$ with a global minimum at $x=0$. Furthermore, we assume that
the walls are completely wet which means that their Young contact angle $\theta=0$ and that the pressure of the bulk reservoir, with which the
confined fluid is in equilibrium, is below the saturated vapour pressure, i.e., $p<p_{\rm sat}$.

At low pressures, the pore is filled with a gas-like phase of a low density $\rho_g$ and the corresponding grand potential per unit length over a
single period can be approximated macroscopically  as
 \bb
\Omega_g=-pS+2\gamma_{\rm wg}\ell_w\,.\label{om_g}
 \ee
Here, $\gamma_{\rm wg}$ is the wall-gas surface tension, $S=PL$ is the area between the walls in the $x$-$z$ plane over one period and
 \bb
 \ell_w=2\int_0^{P/2} \sqrt{1+\psi'^2(x)} \dd x\,, \label{ellw_gen}
 \ee
 is the arc-length of the boundary of each wall in the $x$-$z$ projection over one period.

At sufficiently high pressures, however, the pore will be filled by a liquid-like phase of a high density $\rho_l$, with the grand potential per unit
length
 \bb
 \Omega_l=-p_l S+2\gamma_{\rm wl}\ell_w\,,\label{om_l}
 \ee
where $p_l$ is the pressure of the metastable bulk liquid and $\gamma_{\rm wl}$ is the liquid-wall surface tension. The system undergoes first-order
capillary condensation from the gas-like to the liquid-like phase when $\Omega_g=\Omega_l$. Using Young's equation it follows that the capillary
condensation occurs when the pressure difference $\delta p=p-p_l$ is
 \bb
 \delta p=\frac{2\gamma\ell_w}{S}\,.
 \ee
 More conveniently, this can be expressed in terms of the chemical potential
 \bb
 \delta \mu_{\rm cc}=\frac{2\gamma\ell_w}{S\Delta\rho}\,,\;\;\;({\textrm{gas-liquid}})\,, \label{cc_general}
 \ee
 measuring the shift of the transition from saturation.

Provided the shape of the confining walls is only slowly varying, this can be approximated by
 \bb
 \delta \mu_{\rm cc}=\frac{2\gamma}{L\Delta\rho}\left(1+\delta\right) \label{cc_general2}
 \ee
  where
  \bb
 \delta=\frac{1}{P}\int_{0}^{P/2} \psi'^2(x) \dd x \label{r}
  \ee
is a dimensionless parameter characterizing the wall undulation, which is trivially related with the roughness parameter ($r\approx1+\delta$)
appearing in Eq.~(\ref{wenzel}).

\begin{figure}[h]
\includegraphics[width=0.5\textwidth]{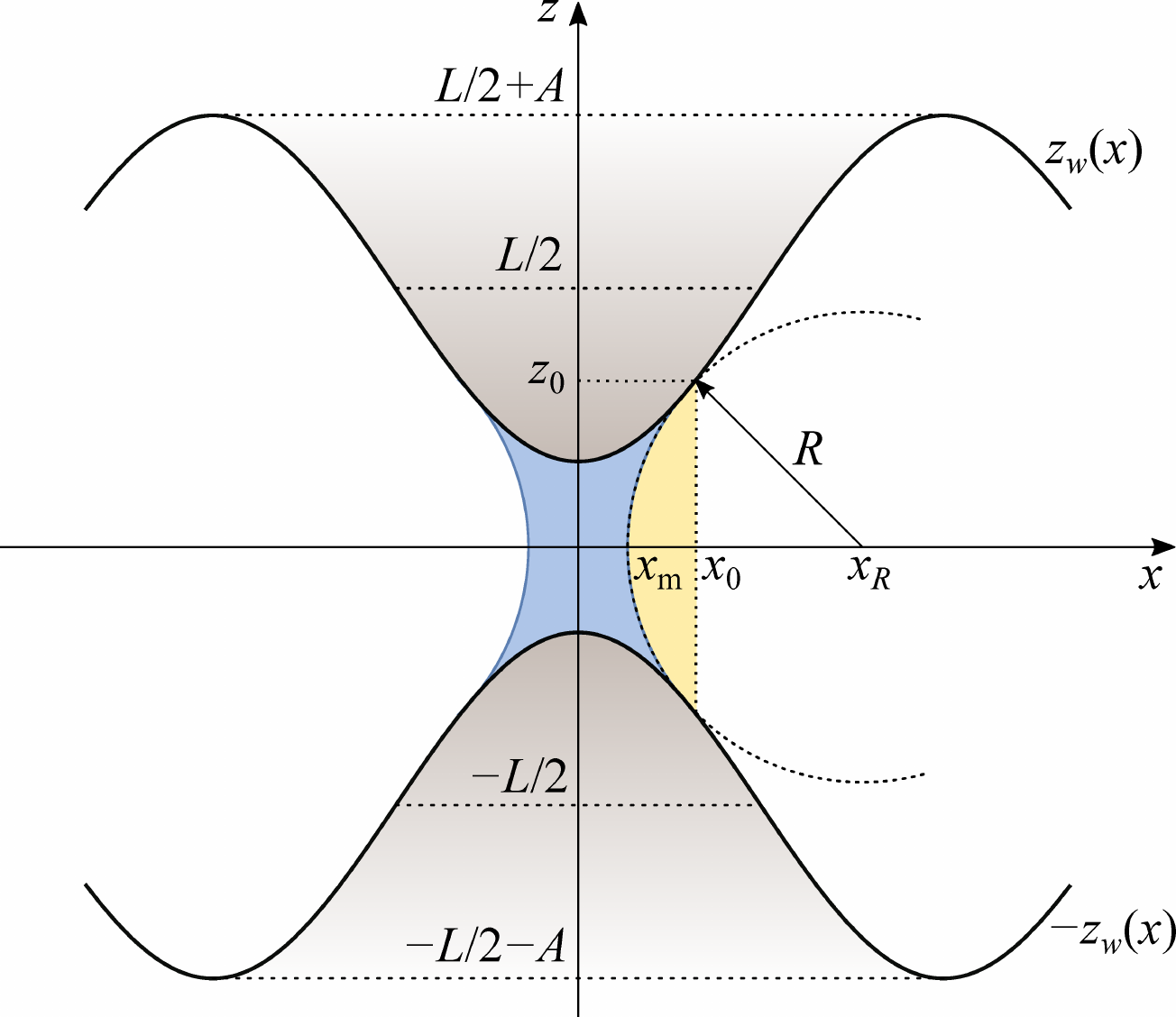}
\caption{A scheme of a bridge phase inside a nanopore formed by two walls of the local height $z_w(x)$ and $-z_w(x)$ relative to the horizontal axis.
The macroscopic picture assumes that the menisci demarcating the liquid bridge are parts of a circle of the Laplace radius $R$ which meets
tangentially the walls at  the points $[\pm x_0,\pm z_0]$.}\label{sketch_complete}
\end{figure}

Apart from the gas-like and the liquid-like phases, the non-planar geometry may also enable a formation of an intermediate phase (or phases), where
the fluid condenses only locally  near the adjacent parts of the walls, giving rise to a periodic array of liquid bridges (see Fig.~2). For
simplicity, we will further assume that the pore geometry allows only for a single bridge per period. The points $[\pm x_0, \pm z_0]$ at which the
menisci of the bridges meet the walls are pressure-dependent and, macroscopically, are specified by two conditions: firstly, the menisci are of a
circular shape with the Laplace radius of curvature $R=\gamma/\delta p$ and, secondly, the menisci meet the walls tangentially  (since the walls are
completely wet). This leads to the implicit equation for $x_0$:
 \bb
 z_w^2(x_0)(1+z_w'^2(x_0))=R^2\,,   \label{x0_impl}
 \ee
 which, together with (\ref{zw_gen}), determines the location of the bridge. This, in turn, allows to obtain a macroscopic approximation for the
 grand potential per unit length of the bridge phase
 \bb
 \Omega_b=-pS_g-p_l S_l+2\gamma_{wg}\ell_w^g+2\gamma_{wl}\ell_w^l+2\gamma\ell\,,\label{om_b}
 \ee
 where
  \bb
  S_l=2\left(2\int_0^{x_0} z_w(x) \dd x-S_m\right)\label{Sl}
  \ee
 and $S_g=S-S_l$, are the volumes (per unit length) occupied by liquid and gas, respectively. Here, the symbol
  \bb
   S_m=R^2\sin^{-1}\left(\frac{z_0}{R}\right)-z_0\sqrt{R^2-z_0^2}
  \ee
  represents the area of to the circular segment highlighted by yellow colour in Fig.~\ref{sketch_complete}.

Furthermore,
 \bb
  \ell_w^l=2\int_0^{x_0}\sqrt{1+\psi'^2(x)}\dd x \label{ellwl}
 \ee
and $\ell_w^g=\ell_w-\ell_w^l$ are the respective arc-lengths of the wall-liquid and wall-gas interfaces. Finally,
 \bb
 \ell=2R\sin^{-1}\left(\frac{z_0}{R}\right) \label{ell}
 \ee
is an arc-length of each meniscus.

First-order bridging transition from G to B occurs at the chemical potential $\mu_{\rm gb}=\mu_{\rm sat}-\delta \mu_{\rm gb}$, when its shift from
saturation is
 \bb
 \delta \mu_{\rm gb}=\frac{2\gamma(\ell_w^l-\ell)}{S_l\Delta\rho}\,,\;\;\;({\textrm{gas-bridge}})\,, \label{gas-bridge}
 \ee
as obtained by balancing $\Omega_g$ and $\Omega_b$. If $\delta\mu_{\rm gb}<\delta\mu_{\rm cc}$, then the bridge state is never the most stable phase
and the bridging transition is preceded by capillary condensation. However, if $\delta\mu_{\rm gb}>\delta\mu_{\rm cc}$, the condensation is a
two-step process, such that the system first condenses locally, when  $\mu=\mu_{\rm gb}$, and eventually globally when $\Omega_b=\Omega_l$, which
occurs for the chemical potential $\mu_{\rm bl}=\mu_{\rm sat}-\delta \mu_{\rm bl}$, with
 \bb
 \delta \mu_{\rm bl}=\frac{2\gamma(\ell+\ell_w^g)}{S_g\Delta\rho}\,,\;\;\;({\textrm{bridge-liquid}})\,. \label{bridge-liquid}
 \ee

\subsection{Sinusoidally shaped walls}

We will now be more specific and consider models of sinusoidally shaped walls by setting
 \bb
   \psi=A\cos(kx)\,,
 \ee
where $A$ is the amplitude and $k=2\pi/P$ is the wave number of the confining walls. In this special case, the geometric measures (5), (\ref{r}),
(\ref{Sl}), and (\ref{ellwl}) become:
 \bb
 \delta=\frac{A^2k^2}{2}\,, \label{r_sin}
 \ee
 \bb
 S_l=2Lx_0-\frac{4A}{k}\sin(kx_0)-2R^2\sin^{-1}\left(\frac{z_0}{R}\right)+2z_0\sqrt{R^2-z_0^2}\,,    \label{Sl_sin}
 \ee
 \bb
 \ell_w^l=2E(x_0,iAk)\,, \label{ellw}
 \ee
 and
  \bb
 \ell_w=\frac{4E(iAk)}{k}\,, \label{l_tot}
 \ee
where $E(\cdot)$ and $E(\cdot,\cdot)$ are the complete and incomplete elliptic integrals of second kind, respectively,  and $i$ is the imaginary
unit.

\subsubsection{Capillary condensation}

It follows from Eqs.~(\ref{cc_general}) and (\ref{l_tot}) that the global condensation from capillary gas to capillary liquid occurs at the chemical
potential:
 \bb
 \delta \mu_{\rm cc}=\frac{4\gamma E(iAk)}{\pi L\Delta\rho}\,, \label{sin_kelvin}
 \ee
which is a simple modification of the Kelvin equation (\ref{kelvin}) for planar slits with completely wet walls ($\theta=0$). This can also be
expressed as a series in the powers of the aspect ratio $a=A/P$:
  \bb
 \delta \mu_{\rm cc}=\delta \mu_{\rm cc}^\parallel\left(1+\pi^2a^2+{\cal{O}}(a^4)\right)\,.\label{sin_kelvin2}
 \ee
From Eq.~(\ref{sin_kelvin2}) it follows that the sinusoidal geometry enhances condensation (as expected), i.e. occurs  farther from saturation
compared to a planar slit. Clearly, this is due to the fact that the area of the (hydrophilic) walls increases with $a$, while the volume of the
metastable liquid in the condensed state remains unchanged. Eq.~(\ref{sin_kelvin2}) also implies that the location of the capillary condensation in
sinusoidal slits does not depend on the wall parameters $A$ and $P$  independently but only on their ratio in a roughly quadratic manner. The
relevance of these macroscopic predictions for microscopic systems will be tested in section \ref{results}.

\subsubsection{Bridging transition}

From Eq.~(\ref{x0_impl}) it follows that the horizontal distance $\pm x_0$ determining the location of the bridge meniscus of radius $R$ is given
implicitly by
  \bb
\left(\frac{L}{2}-A\phi\right)^2\left[1+k^2A^2(1-\phi^2)\right]=R^2\,, \label{phi}
 \ee
with $\phi\equiv\cos(kx_0)$. This is a quartic equation, the solution of which is thus accessible analytically. However, for slightly undulated
walls, $\delta\ll1$, it is more transparent to express $\phi$ as a power series in $\delta$. To this end, we introduce an auxiliary parameter
$\epsilon$:
  \bb
\left(\frac{L}{2}-A\phi\right)^2\left[1+2\epsilon \delta(1-\phi^2)\right]=R^2\,, \label{phi2}
 \ee
 such that the solution is sought in the form of
  \bb
  \phi(\epsilon)=\sum_{n=0}^\infty \phi_n\epsilon^n\,. \label{sum}
  \ee
When plugged into (\ref{phi2}), the coefficients $\phi_n$ are easily determined by balancing the corresponding powers of $\epsilon$:
 \bb
 \phi_0=\frac{\frac{L}{2}-R}{A}\,, \label{phi_0}
 \ee
 \bb
 \phi_1=\frac{Rk^2A(1-\phi_0^2)}{2}\,,
 \ee
etc. Substituting back to (\ref{sum}) and setting $\epsilon=1$, one obtains:
 \bb
 \phi=\frac{L-2R}{2A}+\frac{\delta}{2A}\left(1-\frac{(L-2R)^2}{4A^2}\right)+\mathcal{O}(\delta^2)\,. \label{x0}
 \ee
This can be further simplified by expanding $\phi\approx 1-k^2x_0^2/2$, which to the lowest order in $\delta$ allows for this simple approximation:
 \bb
 x_0\approx\sqrt{\frac{2(1-\phi_0)}{k^2}}\,, \label{x0_0}
 \ee
 with $\phi_0$ given by (\ref{phi_0}).

Once $x_0$ is known, $S_l$ and $\ell_w^l$ (as well as $S_g=S-S_l$ and $\ell_w^g=\ell_w-\ell_w^l$) can be determined from
Eqs.~(\ref{Sl_sin}--\ref{l_tot}). These measures are eventually substituted into Eqs.~(\ref{gas-bridge}) and (\ref{bridge-liquid}) to solve for the
location of the gas-bridge and the bridge-liquid transitions in terms of the corresponding Laplace radii, $R_{\rm gb}=\gamma/(\delta \mu_{\rm
gb}\Delta\rho)$ and $R_{\rm bl}=\gamma/(\delta \mu_{\rm bl}\Delta\rho)$.

 \subsubsection{Spinodals of bridging transitions}

\begin{figure}[h]
\includegraphics[width=0.5\textwidth]{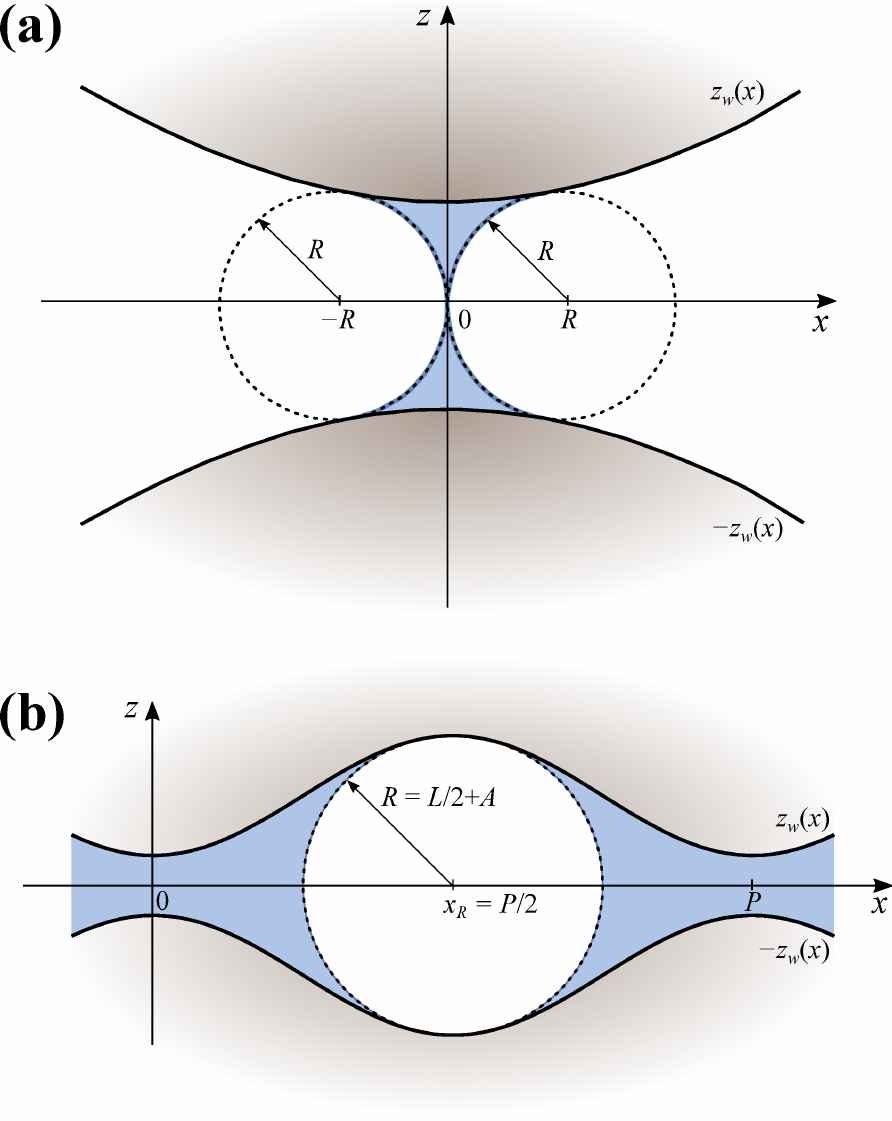}
\caption{Illustration of the macroscopic estimation of the lower (a) and upper (b) spinodals of bridging transition in sinusoidal pores.}\label{spin}
\end{figure}

In contrast to G and L phases, which, on a macroscopic level, have both infinite metastable extensions, the stability of bridging films is restricted
by the pore geometry. As is illustrated in Fig.~\ref{spin}, for the given pore parameters there are lower and upper limits in the values of the
Laplace radius, $R_s^-$ and $R_s^+$, allowing for a formation of the bridging film.

The lower spinodal of B phase corresponds to the smallest Laplace radius, which still enables a formation of the bridge, such that the menisci just
connect each other, cf. Fig.~\ref{spin}a. In order to determine $R_s^-$, we will approximate the shape of the crests by a parabola
 \bb
 z_w(x)\approx c_1+c_2x^2\,, \label{parabola}
 \ee
corresponding to an expansion of $z_w(x)$ to second order around its minimum. Specifically for the sinusoidal pores, the coefficients in
Eq.~(\ref{parabola}) are $c_1=L/2-A$ and $c_2=Ak^2/2$. This approximation seems adequate, since the menisci are close to the origin.

Assuming a circular shape of the menisci, the contact points must satisfy
 \bb
 R_s^-=\frac{x_0^2+z_0^2}{2x_0} \label{spin1}
 \ee
 and the continuity condition further implies that
  \bb
 R_s^-=x_0(2c_2z_0+1)\,. \label{spin2}
  \ee
Eqs.~(\ref{spin1}) and (\ref{spin2}), together with Eq.~(\ref{parabola}) upon substituting for $x_0$, form a set of three equations for three
unknowns, yielding  the contact points of the menisci
   \bb
   x_0=\frac{c_1}{\sqrt{2c_1c_2+1}}\,,  \label{spin3}
   \ee
   \bb
   z_0=c_1\frac{3c_1c_2+1}{2c_1c_2+1}  \label{spin4}
   \ee
   and its radius
   \bb
   R_s^-=\frac{2c_1^2c_2(3c_1c_2+2)}{(2c_1c_2+1)^\frac{3}{2}}\,.  \label{spin5}
   \ee

As for  the largest Laplace radius, $R_s^+$, of a meniscus, which can still fit into the pore, we simply adopt the approximation:
  \bb
  R_s^+=\frac{L}{2}+A\,,  \label{spin6}
  \ee
which corresponds to the state, at which the meniscus meets the walls at the widest part of the pore, see Fig.~\ref{spin}b. This estimation of the
upper spinodal of B phase is justified by the assumption that the aspect ratio $a=A/P$ is not too large.

\section{Mesoscopic corrections} \label{meso}

In this section we extend the macroscopic theory by taking into account the presence of wetting layers adsorbed at the confining walls.

\subsection{Wide pores} \label{meso_wide}

We first consider wide pores experiencing one-step capillary condensation from G to L. In general, the local thickness $\ell(x)$ of wetting layers is
a functional of the wall shape, $\ell(x)=\ell[\psi](x)$, which, in principle, could be contructed using, e.g., a sharp-kink approximation  for
long-range microscopic forces \cite{dietrich} or a non-local interfacial Hamiltonian for short-range microscopic forces \cite{nonlocal}. However,
even for simple wall geometries, such as sinusoids as specifically considered here, either approach would lead to complicated expressions whose
solutions would require numerical treatments. Instead, we propose a simple modification of Derjaguin's correction for the Kelvin equation for planar
slits \cite{evans85, evans85b}. Thus, specifically for long-range microscopic forces and for walls of small roughness, we propose the following
Derjaguin's-like correction for the generalized Kelvin equation (\ref{cc_general})
 \bb
 \delta \mu_{\rm cc}=\frac{2\gamma\ell_w}{(L-3\ell_\pi)\Delta\rho}\,, \label{cc_general_derj}
 \ee
 which for the sinusoidal model becomes
  \bb
 \delta \mu_{\rm cc}=\frac{4\gamma E(iAk)}{\pi (L-3\ell_\pi)\Delta\rho}\,. \label{sin_kelvin_derj}
 \ee
Here, $\ell_\pi$ is the thickness of the wetting layer adsorbed at a single planar wall at the given bulk state. We recall that the factor of $3$ is
associated with the character of the long-range, dispersion forces, which we will consider in our microscopic model and which would be changed to $2$
for short-range forces \cite{evans85, evans85b}.  Clearly, the approximation $\ell(x)\approx\ell_\pi$ seems plausible only for geometries of small
roughness (aspect ratio) which we focus on and for which we will test Eq.~(\ref{sin_kelvin_derj}) by comparing with DFT results. Furthermore, taking
into account that Eqs.~(\ref{cc_general_derj}) and (\ref{sin_kelvin_derj}) refer to wide pores, capillary condensation is expected to occur near the
bulk coexistence where $\ell_\pi$ can be described analytically in its known asymptotic form.

\subsection{Narrow pores} \label{meso_narrow}

\begin{figure}[h!]
\includegraphics[width=0.45\textwidth]{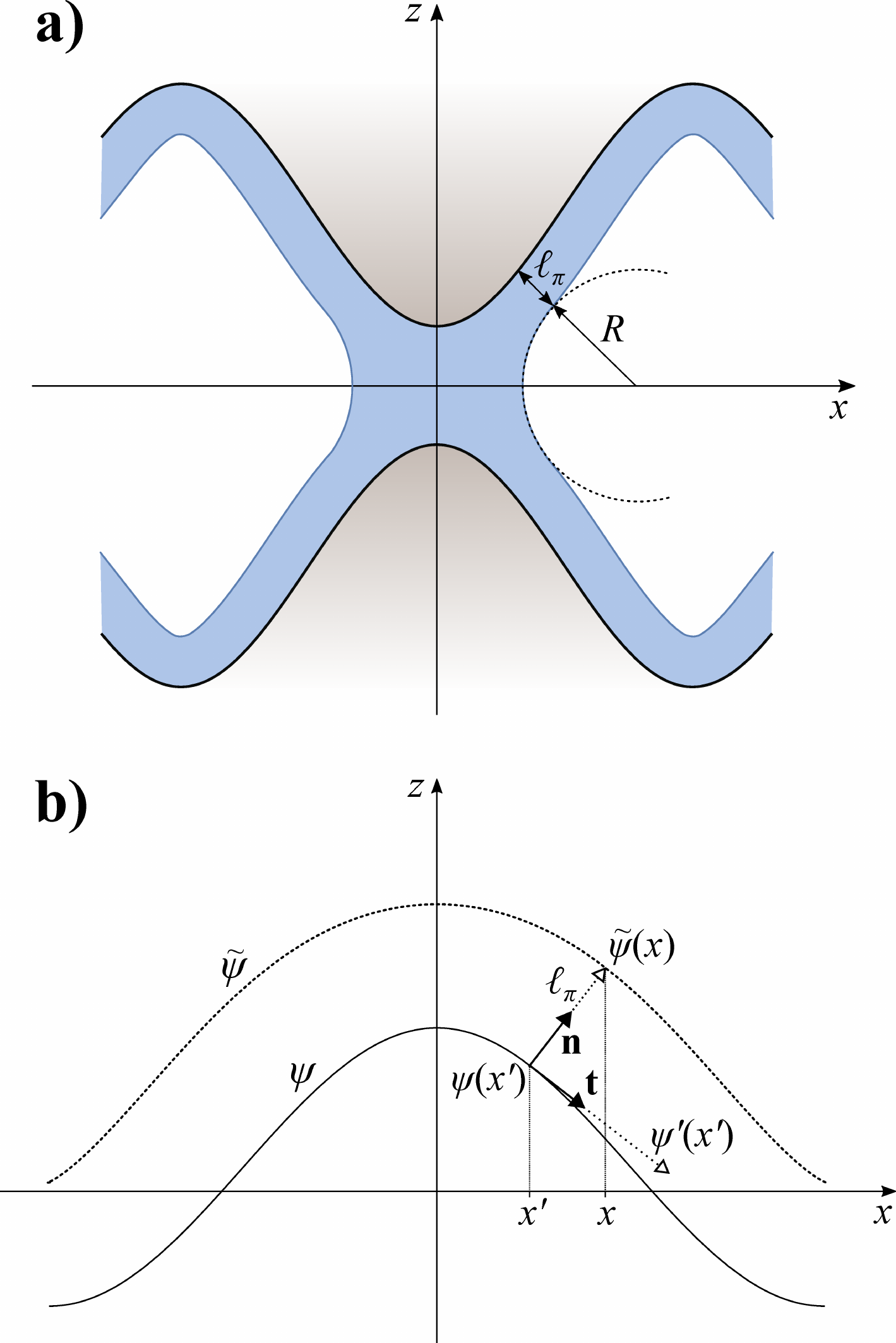}
\caption{Illustration of the RP geometric construction of the bridge phase in a completely wet sinusoidal nanopore by taking into account the wetting
layers \cite{nature}. a) The walls are first coated by wetting layers, whose normal width is $\ell_\pi$ corresponding to a thickness of the liquid
film adsorbed on a planar wall at the given chemical potential. In the second step, the circular menisci of the Laplace radius $R=\gamma/(\delta
\mu\Delta\rho)$ are drawn, such that they meet the wetting layers tangentially; b) The construction of the shape of the interface $\tilde{\psi}(x)$
corresponding to the wetting layers. The unit vectors $\mathbf{n}$ and $\mathbf{t}$ are normal and tangent to the wall at a given point $x'$,
respectively, using which the height of the wetting layer can be determined at the point $x$, shifted from $x'$ according to
Eq.~(\ref{shift_x}).}\label{RP}
\end{figure}

For narrow pores of widths $L<L_t$, condensation occurs via formation of capillary bridges. To account for wetting layers in this case, we will adopt
the geometric construction due to Rasc\'on and Parry (RP) \cite{nature}, which is schematically illustrated in Fig.~\ref{RP}a. The construction
consists of two steps: i) first, each wall is covered by a wetting film whose width measured normally to the wall is $\ell_\pi$, ii) secondly,
menisci of the Laplace radius $R=\gamma/(\delta \mu\Delta\rho)$ are connected tangentially to the \emph{wetting layers} (rather than to the walls).
By following this rule, we will first show explicitly what the shape $\tilde\psi(x)$ of the wetting film interface is for a general shape $\psi(x)$
of the wall, before applying this result specifically for the sinusoidal wall.

Let us consider an arbitrary point $x'$ on the horizontal axis, at which the local height of the wall is $\psi(x')$. Thus, the unit tangential vector
at this point is ${\mathbf t}=(1,\psi'(x'))/\sqrt{1+\psi'^2(x')}$, where the prime denotes a derivative with respect to $x'$; hence, the unit normal
at $\psi(x')$ is ${\mathbf n}=(-\psi'(x'),1)/\sqrt{1+\psi'^2(x')}$. According to the RP construction the local height of the wetting film interface
$\tilde{\psi}(x)$ is a distance $\ell_\pi$ from $\psi(x')$ along the normal vector, (see Fig.~\ref{RP}b). It follows that
 \bb
  \tilde{\psi}(x)=\psi(x')+\frac{\ell_\pi}{\sqrt{1+\psi'^2(x')}} \label{shift_z}
 \ee
 where
 \bb
 x=x'-\frac{\ell_\pi\psi'(x')}{\sqrt{1+\psi'^2(x')}} \label{shift_x}\,.
 \ee
 Considering walls of small gradients, the difference $x-x'$ is supposed to be small, thus
  \bb
  \tilde{\psi}(x)\approx\psi(x')+\frac{\ell_\pi}{\sqrt{1+\psi'^2(x)}} \label{shift_z2}
 \ee
 and
  \bb
 x'\approx x+\frac{\ell_\pi\psi'(x)}{\sqrt{1+\psi'^2(x)}} \label{shift_x2}\,,
 \ee
to first order in $x-x'$. By substituting (\ref{shift_x2}) into (\ref{shift_z2}), one obtains that
 \bb
  \tilde{\psi}(x)\approx\psi\left(x+\frac{\ell_\pi\psi'(x)}{\sqrt{1+\psi'^2(x)}}\right)+\frac{\ell_\pi}{\sqrt{1+\psi'^2(x)}}\,, \label{shift_fin}
 \ee
 which determines $\tilde{\psi}(x)$ explicitly. This can be further simplified by expanding the first term on the r.h.s. to first order:
 \bb
  \tilde{\psi}(x)\approx \psi(x)+\ell_\pi\sqrt{1+\psi'^2(x)}\,. \label{shift_fin}
 \ee
Specifically, for the sinusoidal wall, Eq.~(\ref{shift_fin}) becomes:
 \bb
 \tilde{\psi}(x)\approx A\cos(kx)+\ell_\pi\sqrt{1+A^2k^2\sin^2(kx)}\,. \label{rp_sin}
 \ee

Thus, within the mesoscopic treatment, we proceed in the same manner as in the previous section, except that $\psi(x)$ is replaced by
$\tilde{\psi}(x)$, as given by Eq.~(\ref{rp_sin}).

\section{Density functional theory} \label{dft}

Classical DFT \cite{evans79} is a tool of statistical mechanics describing equilibrium behaviour of inhomogeneous fluids.  Based on the variational
principle, the equilibrium one-body density $\rhor$ of the fluid particles is determined by  minimizing the grand potential functional:
 \bb
 \Omega[\rho]={\cal F}[\rho]+\int\dd\rr\rhor[V(\rr)-\mu]\,. \label{grandpot}
 \ee
Here, ${\cal F}[\rho]$ is the intrinsic free-energy functional, which contains all the information about the intermolecular interactions between the
fluid particles, $V(\rr)$ is the external potential, which, in our case, represents the influence of the confining walls and $\mu$ is the chemical
potential of the system and the bulk reservoir. The intrinsic free-energy functional is usually separated into two parts:
 \bb
 {\cal F}[\rho]={\cal F}_{\rm id}[\rho]+{\cal F}_{\rm ex}[\rho]\,.
 \ee
The first, ideal-gas contribution, which is due to purely entropic effects, is known exactly:
     \bb
  \beta {\cal F}_{\rm id}[\rho]=\int\dr\rho(\rr)\left[\ln(\rhor\Lambda^3)-1\right]\,,
  \ee
where $\Lambda$ is the thermal de Broglie wavelength and $\beta=1/k_BT$ is the inverse temperature.

The remaining excess part of the intrinsic free energy arising from the fluid-fluid interaction, ${\cal F}_{\rm ex}$, must be almost always
approximated and its  treatment depends on the interaction model. For models involving hard cores, the excess contribution can be treated in a
perturbative manner, such that it is typically further split into the contribution ${\cal F}_{\rm hs}$ due to hard-sphere repulsion, and the
contribution ${\cal F}_{\rm att}$ arising from attractive interactions:
 \bb
{\cal F}_{\rm ex}[\rho]={\cal F}_{\rm hs}[\rho]+{\cal F}_{\rm att}[\rho]\,.
 \ee
The hard-sphere part of the free-energy is described using Rosenfeld's fundamental measure theory  \cite{ros}
 \bb
{\cal F}_{\rm hs}[\rho]=k_BT\int\dd\rr\,\Phi(\{n_\alpha\})\,, \label{fhs}
 \ee
where  the free energy density $\Phi$ depends on the set of weighted densities $\{n_\alpha\}$. Within the original Rosenfeld approach these consist
of four scalar and two vector functions, which are given by convolutions of the density profile and the corresponding weight function:
 \bb
 n_\alpha(\rr)=\int\dr'\rho(\rr')w_\alpha(\rr-\rr')\;\;\alpha=\{0,1,2,3,v1,v2\}\,,  \label{weightden}
 \ee
where $w_3(\rr)=\Theta(R-|\rr|)$, $w_2(\rr)=\delta(R-|\rr|)$, $w_1(\rr)=w_2(\rr)/4\pi R$, $w_0(\rr)=w_2(\rr)/4\pi R^2$,
$w_{v2}(\rr)=\rr\delta(R-|\rr|)/R$, and $w_{v1}(\rr)=w_{v2}(\rr)/4\pi R$. Here, $\Theta$ is the Heaviside function, $\delta$ is the Dirac function
and $R=\sigma/2$ where $\sigma$ is the hard-sphere diameter.

The attractive free-energy contribution is treated at a mean-field level:
 \bb
 F_{\rm att}[\rho]=\frac{1}{2}\int d{\bf{r}}_1\rho(\rr_1)\int d{\bf{r}}_2\rho(\rr_2)u_{\rm att}(|\rr_1-\rr_2|)\,, \label{fatt}
 \ee
 where $u_{\rm att}(r)$ is the attractive part of the Lennard-Jones-like potential:
 \bb
 u_{\rm att}(r)=\left\{\begin{array}{cc}
 0\,;&r<\sigma\,,\\
-4\varepsilon\left(\frac{\sigma}{r}\right)^6\,;& \sigma<r<r_c\,,\\
0\,;&r>r_c\,.
\end{array}\right.\label{ua}
 \ee
which is truncated at $r_c=2.5\,\sigma$. For this model, the critical temperature corresponds to $k_BT_c=1.41\,\varepsilon$.

The external potential $V(\rr)=V(x,z)$ representing the presence of the confining walls can be expressed as follows:
 \bb
 V(x,z)=V_w(x,L/2+z)+V_w(x,L/2-z)\,,
 \ee
where $L$ is the mean distance between the walls and $V_w(x,z)$ describes a potential of a single, sinusoidally shaped wall with an amplitude $A$ and
period $P=2\pi/k$, formed by the Lennard-Jones atoms distributed uniformly with a density $\rho_w$:
 \begin{eqnarray}
 V_w(x,z)&=&\rho_w\int_{-\infty}^\infty\dd x' \int_{-\infty}^\infty\dd y'\int_{-\infty}^{A\cos(kx')}\dd z'\nonumber\\
   &&u_w \left(\sqrt{(x-x')^2+y'^2+(z-z')^2} \right)\,,  \label{wallpot}
 \end{eqnarray}
where
\begin{equation}
  u_w(r) =4\,\varepsilon_w\left[\left(\frac{\sigma_w}{r}\right)^{12}-\left(\frac{\sigma_w}{r}\right)^{6}\right]
\end{equation}
is  the 12-6 Lennard-Jones potential.

Minimization of \eqref{grandpot} leads to the Euler--Lagrange equation
  \bb
  \label{el}
  \frac{\delta{\cal F}[\rho]}{\delta\rho(\rr)} + V(\rr) - \mu = 0\,,
 \ee
which can be recast into the form of a self-consistent equation for the equilibrium density profile:
 \bb
  \label{selfconsistent}
  \rho(\rr) = \Lambda^{-3} \exp\left[\beta\mu-\beta V(\rr) + c^{(1)}(\rr)\right]
 \ee
 that can be solved iteratively. Here, $c^{(1)}(\rr)=c^{(1)}_\mathrm{hs}(\rr)+c^{(1)}_\mathrm{att}(\rr)$ is the one-body direct correlation
function, whose hard-sphere contribution,
  \bb
  \label{chs}
  c^{(1)}_{\rm hs}(\rr)= -\sum_\alpha \int\dd\rr'\; \frac{\partial\Phi(\{n_\alpha\})}{\partial n_\alpha} \, w_\alpha(\rr'-\rr)
 \ee
 and the attractive contribution,
  \bb
  \label{catt}
  c^{(1)}_\mathrm{att}(\rr) = -\beta\int \dd\rr'\; u_{\rm att}(|\rr-\rr'|)\,\rho(\rr'),
 \ee
  are obtained  by varying ${\cal F}_{\rm hs}$ and ${\cal F}_{\rm att}$ w.r.t. $\rho(\rr)$, respectively.

Eq.~\eqref{selfconsistent} was solved numerically using Picard's iteration  on a 2D rectangular grid with an equidistant spacing of $0.1\,\sigma$
(except for the calculations presented in Fig.~\ref{cc_fix_a}, where the considered wall parameters required reducing of the grid spacing down to
$0.02\,\sigma$). For evaluations of the integrals \eqref{weightden}, \eqref{chs}, and \eqref{catt}, which are in the form of convolutions, we applied
the Fourier transform. To this end, we followed the approach of Salinger and Frink \cite{frink2003},  according to which Fourier transforms of
$\rho(\rr)$ and $\partial\Phi(\{n_\alpha\}) /
\partial n_\alpha$ are evaluated numerically using the fast Fourier transform, while $\hat w_\alpha$ are calculated analytically \cite{four_conv}:
\begin{eqnarray*}
&&\hat w_3(\kk)=4\pi R^3\,\frac{\sin(2\pi kR)-2\pi k \cos(2\pi kR)}{(2\pi k R)^3}, \\
&&\hat w_2(\kk)=4\pi R^2\,\frac{\sin(2\pi kR)}{2\pi k R},\\
&&\hat w_1(\kk)=\frac{\hat w_2(\kk)}{4\pi R},\hspace*{1.75cm}
\hat w_0(\kk)=\frac{\hat w_2(\kk)}{4\pi R^2},\\
&&\hat w_\mathrm{v2}(\kk)=-2\pi \kk \hat w_3(\kk),\qquad \hat w_\mathrm{v1}(\kk)=\frac{\hat w_\mathrm{v2}(\kk)}{4\pi R}\,,
\end{eqnarray*}
where $\kk=(k_x,k_z)$ is the vector in the reciprocal space and $k=|\kk|$. We applied the analogous approach to evaluate the attractive contribution
to the one-body direct correlation function, $c^{(1)}_\mathrm{att}(\rr)$, as given by Eq.~\eqref{catt}. To this end, the Fourier transform of $u_{\rm
att}(r)$ has been determined analytically:
  \bb
  \label{ftuatt}
  \hat u_\mathrm{att}(k) = \frac{2\,\varepsilon\sigma^2}{3\,k r_c^4}
    \left[r_c^4\, \Psi(k;\sigma) - \sigma^4\, \Psi(k;r_c)\right]\,,
 \ee
 where
  \begin{eqnarray}
  \Psi(k;\xi) &=&  2\pi k\xi\left(2\pi^2 k^2\xi^2-1\right) \cos\left(2\pi k\xi\right)\nonumber \\
                        &&+ \left(2\pi^2 k^2\xi^2-3\right) \sin\left(2\pi k\xi\right)\\
                        &&+ 8\pi^4 k^4 \xi^4 \Si\left(2\pi k \xi \right)\,, \nonumber
    \end{eqnarray}
where $\Si(x)=\int_0^x \sin(t)/t\dd t$ is the sine integral.

Once the equilibrium density is obtained, the phase behaviour of the system can be studied by determining the grand potential, as given by
substituting $\rhor$ back to \eqref{grandpot}, and the adsorption, defined as
  \bb
  \Gamma = \frac{1}{LP}\int_0^P \dd x \int_{-z_w(x)}^{z_w(x)} \dd z \; \left[\rho(x,z)-\rho_b \right]\,,
 \ee
where $\rho_b$ is the density of the bulk gas.

\section{Results and discussion} \label{results}

\begin{figure*}[t]
\includegraphics[width=0.32\textwidth]{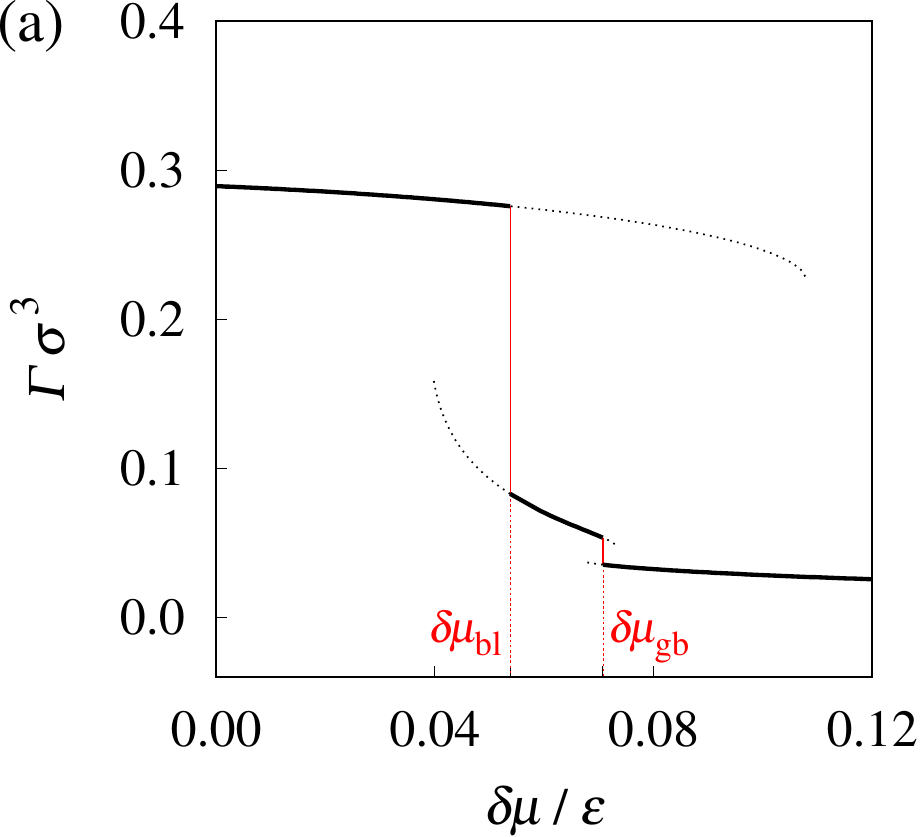} \includegraphics[width=0.32\textwidth]{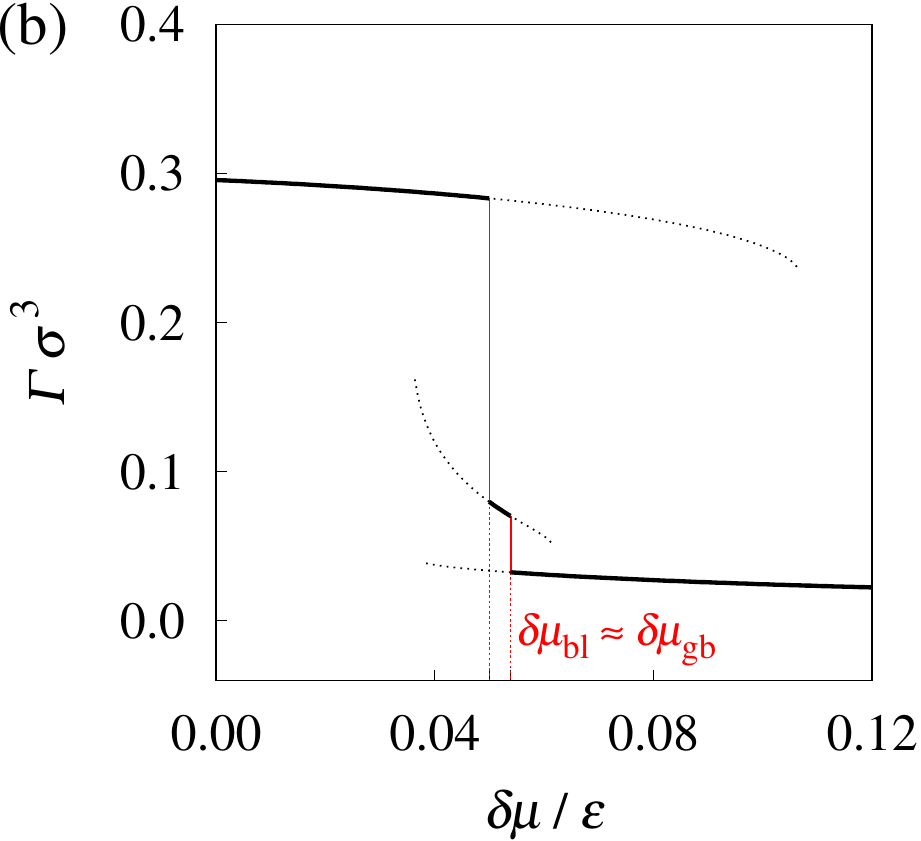} \includegraphics[width=0.32\textwidth]{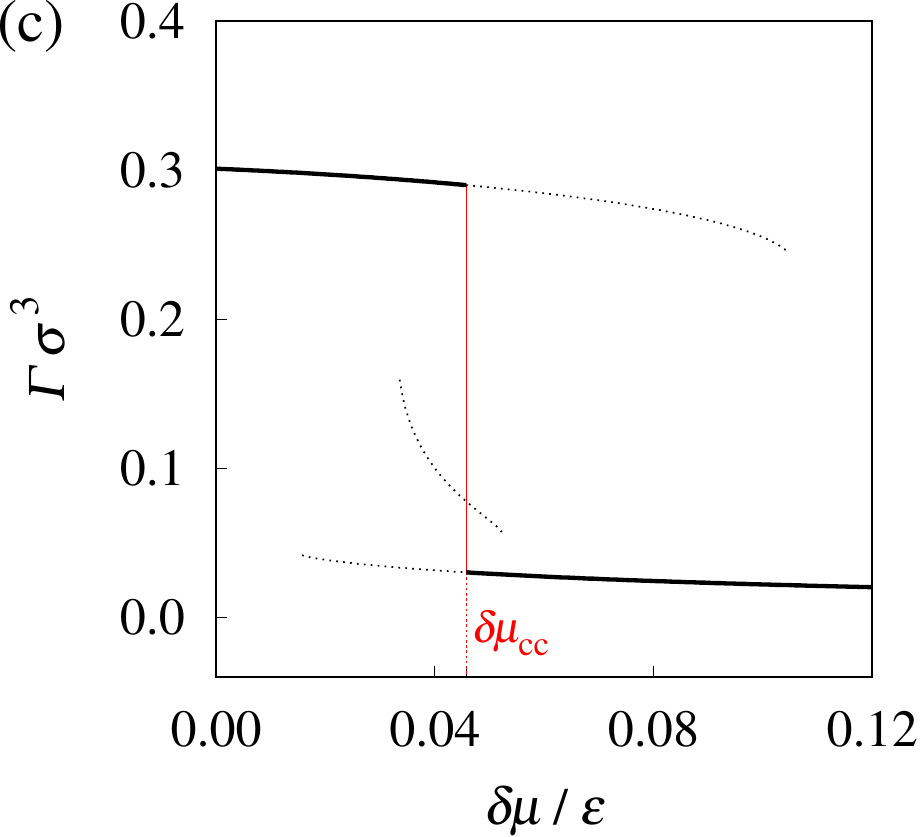}
\caption{Adsorption isotherms obtained from DFT for nanopores formed by walls with $A=2\,\sigma$ and $P=50\,\sigma$. The mean distance  between the
walls is a) $L=8\,\sigma$, b) $L=9\,\sigma$, and c) $L=10\,\sigma$.}\label{ads}
\end{figure*}

In this section, we present our DFT results for condensation of simple fluids confined by two sinusoidally shaped walls using the model presented in
the previous section for the wall parameters $\varepsilon_w=0.8\,\varepsilon$ and $\sigma_w=\sigma$. The results are compared with the predictions
based on the macroscopic and mesoscopic arguments formulated in sections \ref{macro} and \ref{meso}. In order to test the quality of the predictions,
we will consider two temperatures. We will first present our results for temperature $k_B T/\varepsilon\doteq1.28\approx k_BT_w/\varepsilon$, which
is slightly \emph{below} the wetting temperature. At this temperature, the contact angle of the considered walls is very low (about $1^\circ$), which
means that macroscopically the walls can be viewed effectively as completely wet, yet they remain covered by only a microscopically thin wetting
films (since the isolated walls exhibit first-order wetting). The reason behind this choice is that we, first of all, wish to test the quality of the
purely macroscopic theory, which ignores the presence of wetting layers adsorbed at the walls. Clearly, if the theory did not work reasonably well
even in the absence of wetting layers, then any attempt of its elaboration by including mesoscopic corrections accounting for the presence of wetting
layers would not be meaningful. However, we will show that the macroscopic theory is in a close agreement with the DFT results for all the types of
phase transitions the system experiences, and provides thus quantitatively accurate description of the phase diagrams for the considered nanopores.
In the next step, we will consider a higher temperature, $k_B T/\varepsilon=1.35$, which is well above the wetting temperature, and compare the DFT
results with both the purely macroscopic theory, as well as its mesoscopic modification. If not stated otherwise, the comparison will be illustrated
by considering walls with a period $P=50\,\sigma$ and amplitudes $A=2\,\sigma$ or $A=5\,\sigma$. We deliberately avoid systems with large aspect
ratios for the reason discussed in the concluding section.


\subsection{$T\approx T_w$}

\begin{figure}[t]
\includegraphics[width=0.45\textwidth]{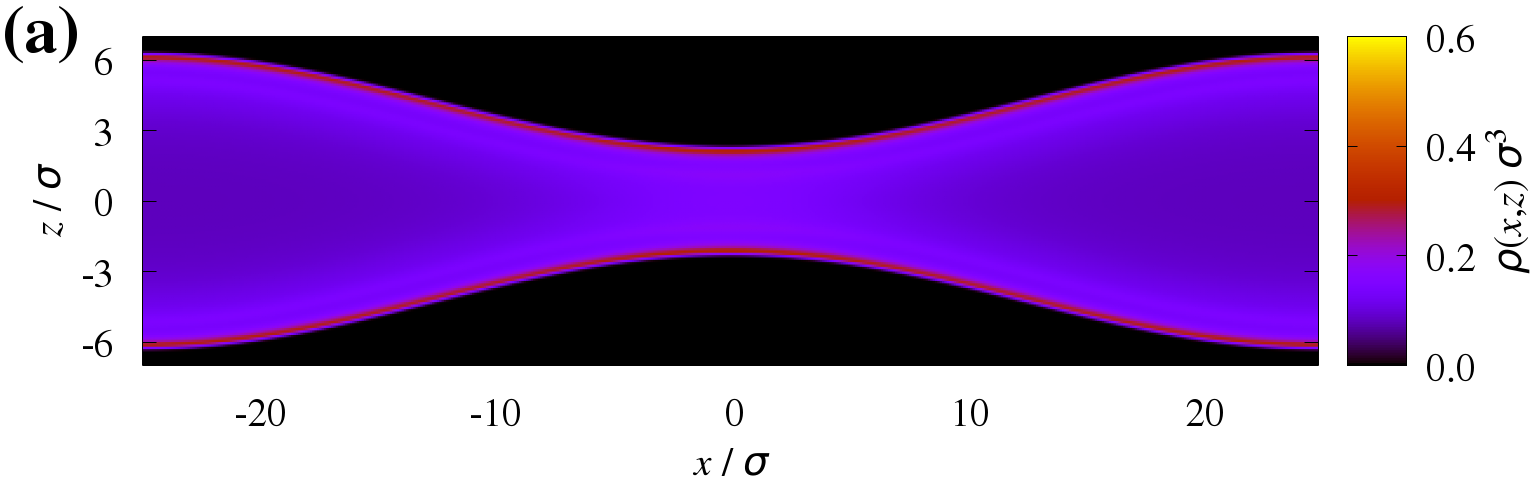}
\includegraphics[width=0.45\textwidth]{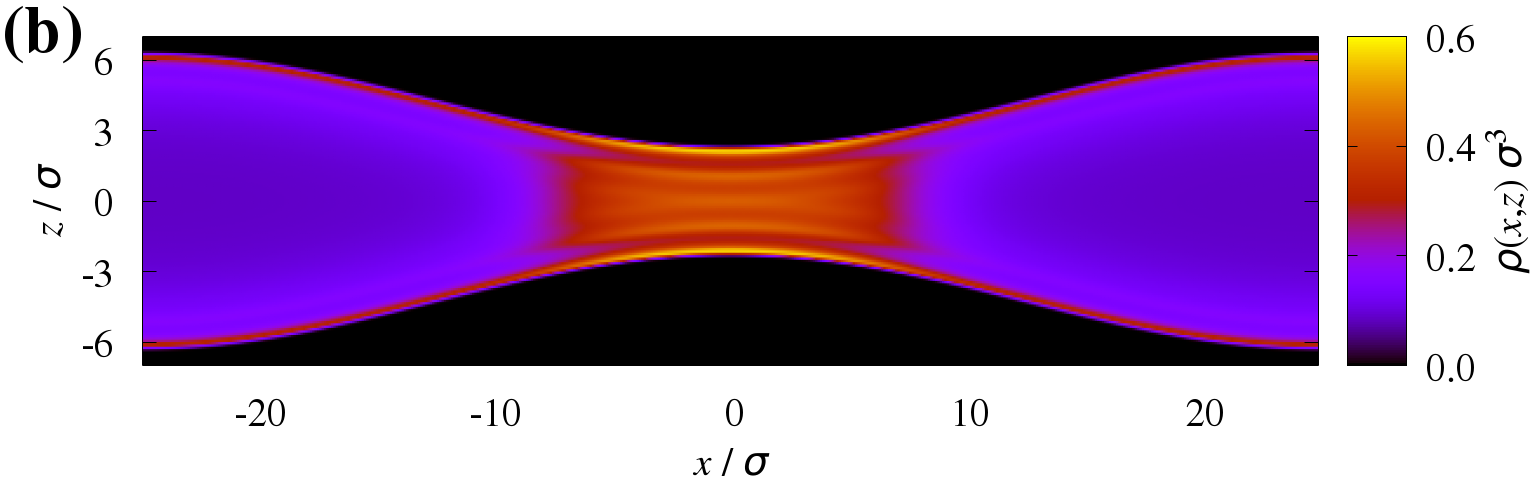}
\includegraphics[width=0.45\textwidth]{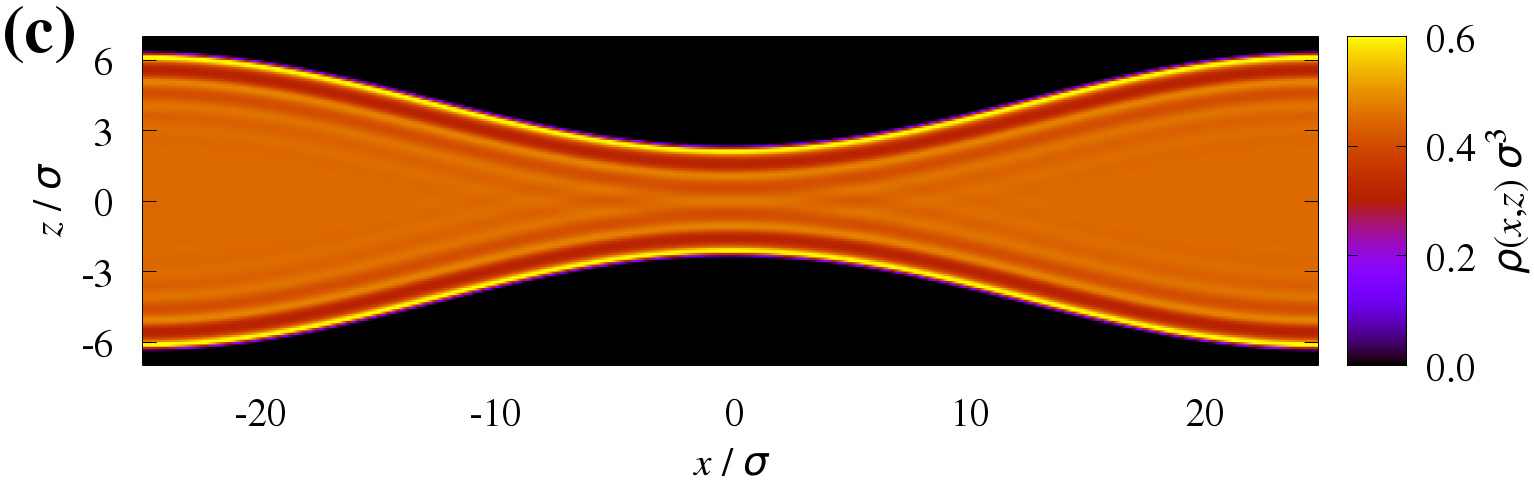}
 \caption{Equilibrium 2D density profiles corresponding to a) capillary gas, b) bridge, and c) capillary liquid phases in the nanopore
with $A=2\,\sigma$, $P=50\,\sigma$ and $L=8\,\sigma$ (cf. Fig.~\ref{ads}a).}\label{dens_profs_A2_L8}
\end{figure}

We start with presenting adsorption isotherms obtained from DFT for nanopores with fixed wall parameters but for different mean widths $L$ (see Fig.~\ref{ads}). For the
smallest $L$, the adsorption isotherm exhibits two jumps separating three capillary phases. As expected, these correspond to G, which is stable sufficiently far from
saturation, B which is stabilized at intermediate pressures and L, which forms close to saturation. The structure of all the capillary phases are illustrated in
Fig.~\ref{dens_profs_A2_L8} where the 2D equilibrium density profiles are plotted. As the mean width of the pore $L$ is increased, the interval of $\delta\mu$ over which
the bridge phase is stable becomes smaller and smaller, as is illustrated in Fig.~\ref{ads}b . Here, the locations of G-B and B-L transitions become almost identical,
which means that such a value of $L$ is already very close to $L_t$ allowing for G-B-L coexistence.  For  $L>L_t$, the bridge phase is never the most stable state, so
that capillary gas condenses to capillary liquid directly in a single-step (cf. Fig.~\ref{ads}c).

\begin{figure}[h]
\includegraphics[width=0.45\textwidth]{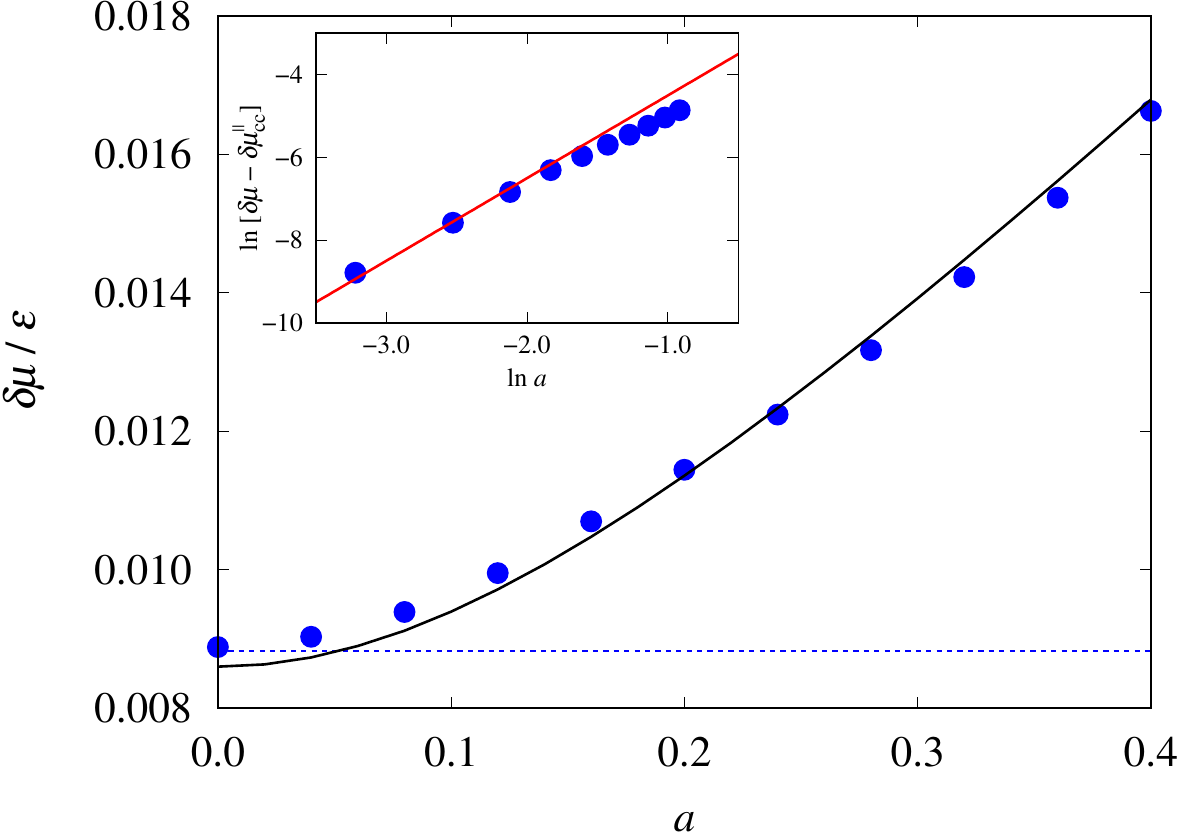}
\caption{DFT results (symbols) showing a dependence of $\delta\mu_{\rm cc}$ on the aspect ratio $a=A/P$ for nanopores with $P=50\,\sigma$ and
$L=50\,\sigma$. The solid line represents the solution of the Kelvin equation (\ref{sin_kelvin2}) and the dashed line shows the value of
$\delta\mu_{\rm cc}^\parallel$ for capillary condensation in the planar slit obtained from 1D DFT. The inset shows the log-log plot of the DFT
results and the straight line with the slope of $2$ confirms the prediction \eqref{sin_kelvin2}. }\label{cc_vary_a}
\end{figure}

\begin{figure}[h]
\includegraphics[width=0.45\textwidth]{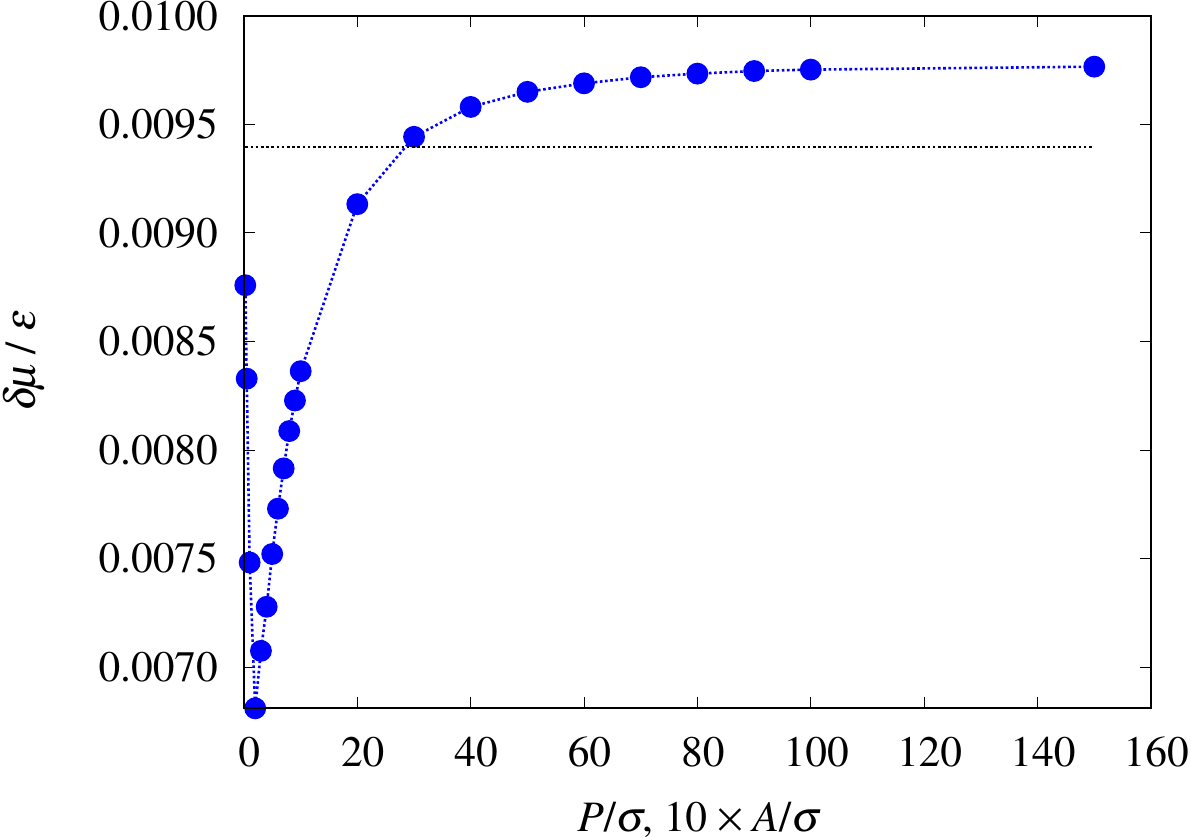}
\caption{DFT results for a dependence of $\delta\mu_{\rm cc}$ on $A$ and $P$, such that $a=A/P=0.1$.  The horizontal dotted line indicates the
prediction given by Kelvin's equation (\ref{sin_kelvin}). }\label{cc_fix_a}
\end{figure}

\begin{figure}[h]
\includegraphics[width=0.45\textwidth]{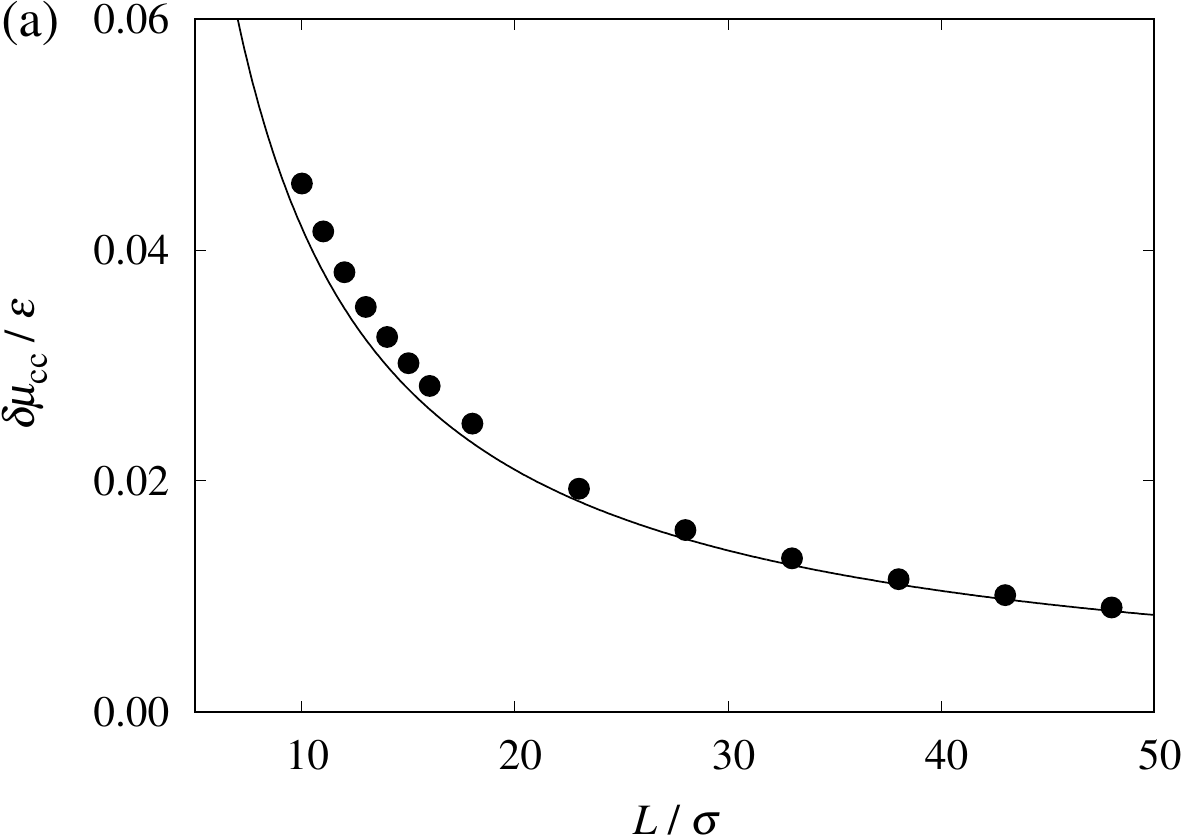}
\includegraphics[width=0.45\textwidth]{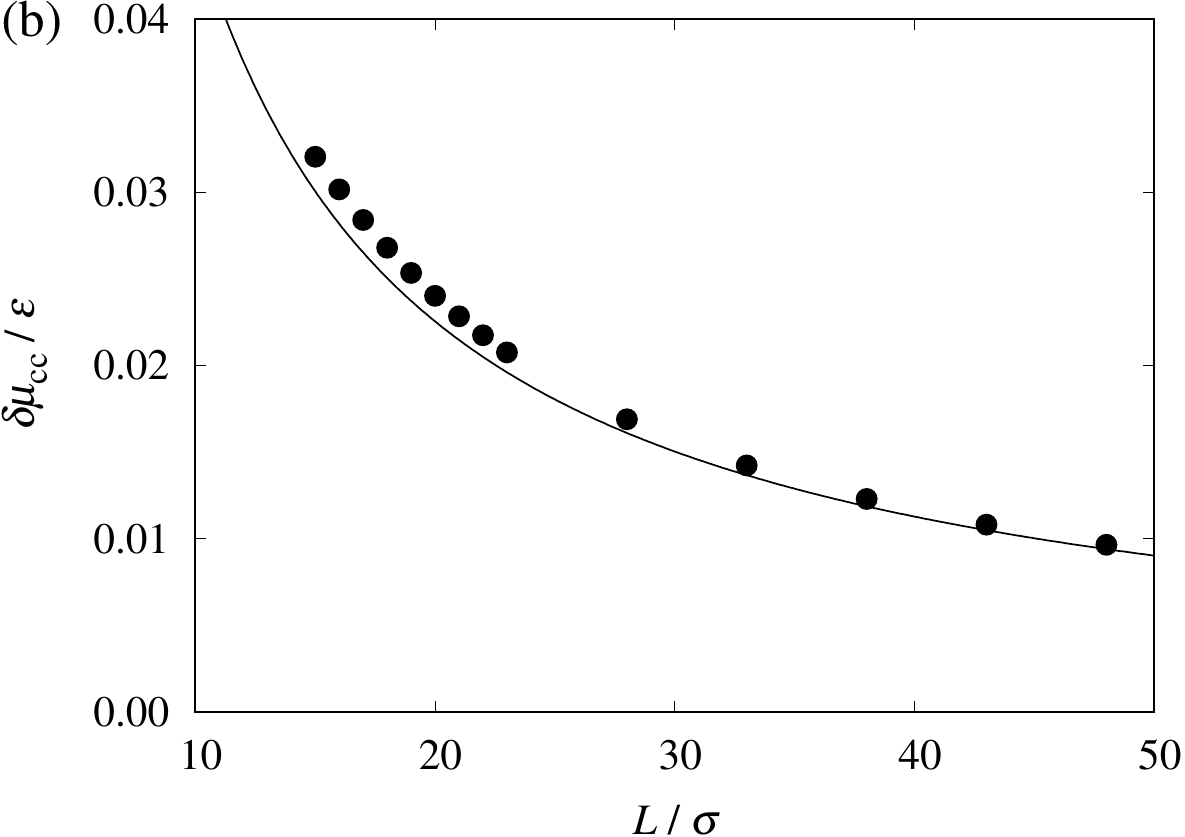}
\caption{A comparison between DFT results (symbols) and the prediction given by Kelvin's equation (\ref{sin_kelvin}) (line) for a dependence of
$\delta\mu_{\rm cc}$ on $L$ for walls with amplitudes $A=2\,\sigma$ (a) and $A=5\,\sigma$ (b) and period $P=50\,\sigma$.}\label{cc_vary_L}
\end{figure}

Let us first focus on a single-step capillary condensation at wide slits. Fig.~\ref{cc_vary_a} displays DFT results showing a  dependence of
$\delta\mu_{\rm cc}$ on the wall amplitude up to $A\approx20\,\sigma$ (with both $P$ and $L$ fixed to $50\,\sigma$). The agreement between DFT
results and the Kelvin equation (\ref{sin_kelvin}) is very good, and in particular the inset Fig.~\ref{cc_vary_a} confirms that the dependence
$\delta\mu_{\rm cc}(a)$ is approximately quadratic for sufficiently small amplitudes, in line with the expansion  (\ref{sin_kelvin2}). We note that
the results include the case of $A=0$ corresponding to a planar slit (in which case the walls exert the standard $9$-$3$ Lennard-Jones potential),
obtained independently using 2D, as well as a simple 1D DFT; the resulting values of $\delta\mu_{\rm cc}$ are essentially identical, which serves as
a good test of the numerics.

\begin{figure}[h]
\includegraphics[width=0.45\textwidth]{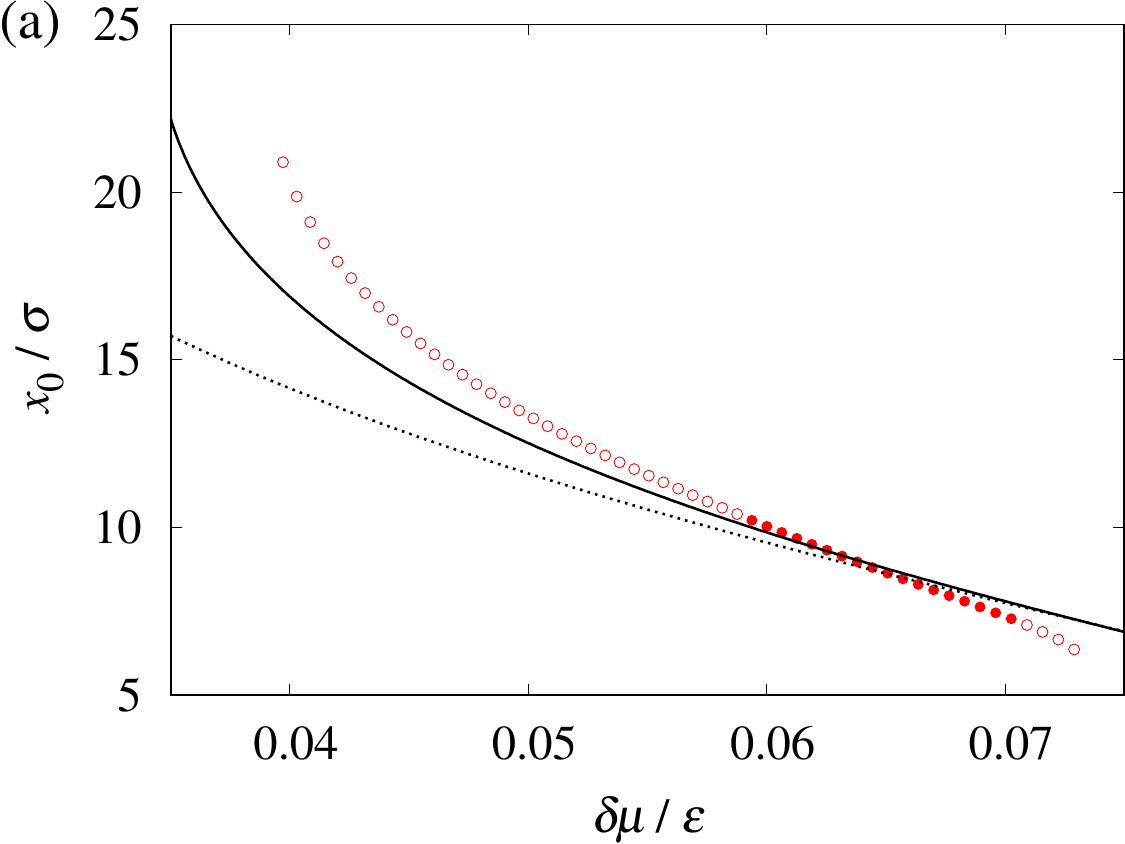}
\includegraphics[width=0.45\textwidth]{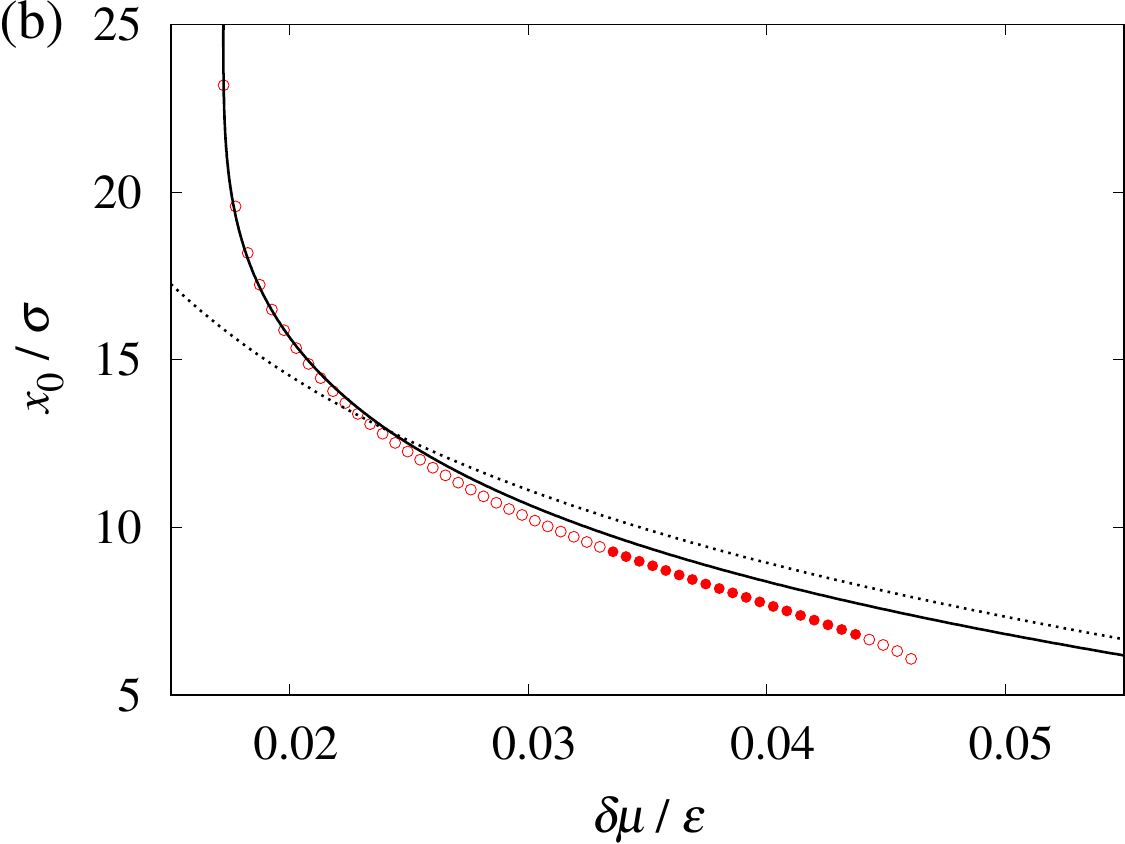}
\caption{A dependence of $x_0$, specifying the location where the bridging menisci meet the walls, on $\delta\mu$, for the slits with $A=2\,\sigma$
and $L=8\,\sigma$ (a) and $A=5\,\sigma$ and $L=14\,\sigma$ (b). The period of the walls is $P=50\,\sigma$ in both cases. A comparison is made between
DFT results (symbols), the prediction given by the solution of the quartic equation, \eqref{phi}, (full line)) and its simple approximative solution,
\eqref{x0_0}, based on the perturbative scheme (dotted line)). The DFT results include states where the bridges are stable (full circles), as well as
the states where the bridges are metastable (open circles). }\label{x0_A2A5}
\end{figure}

\begin{figure}[h]
\includegraphics[width=0.45\textwidth]{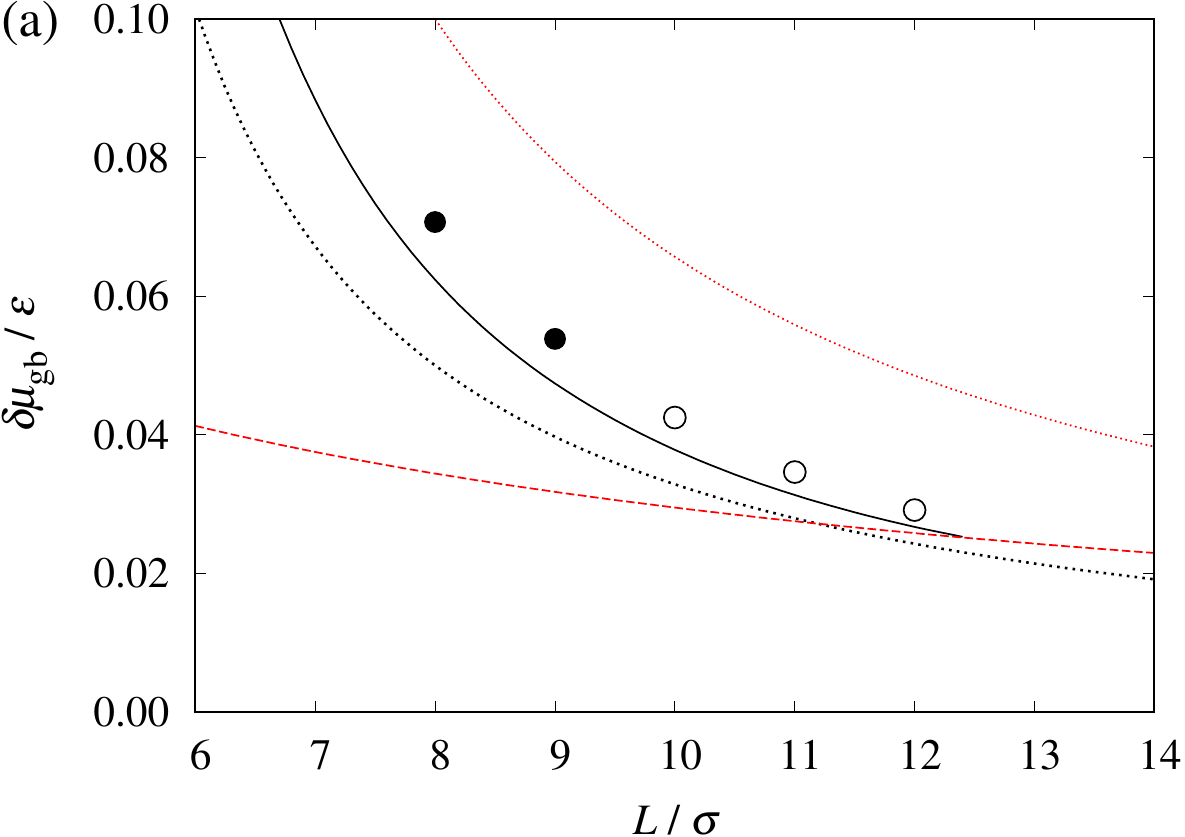}
\includegraphics[width=0.45\textwidth]{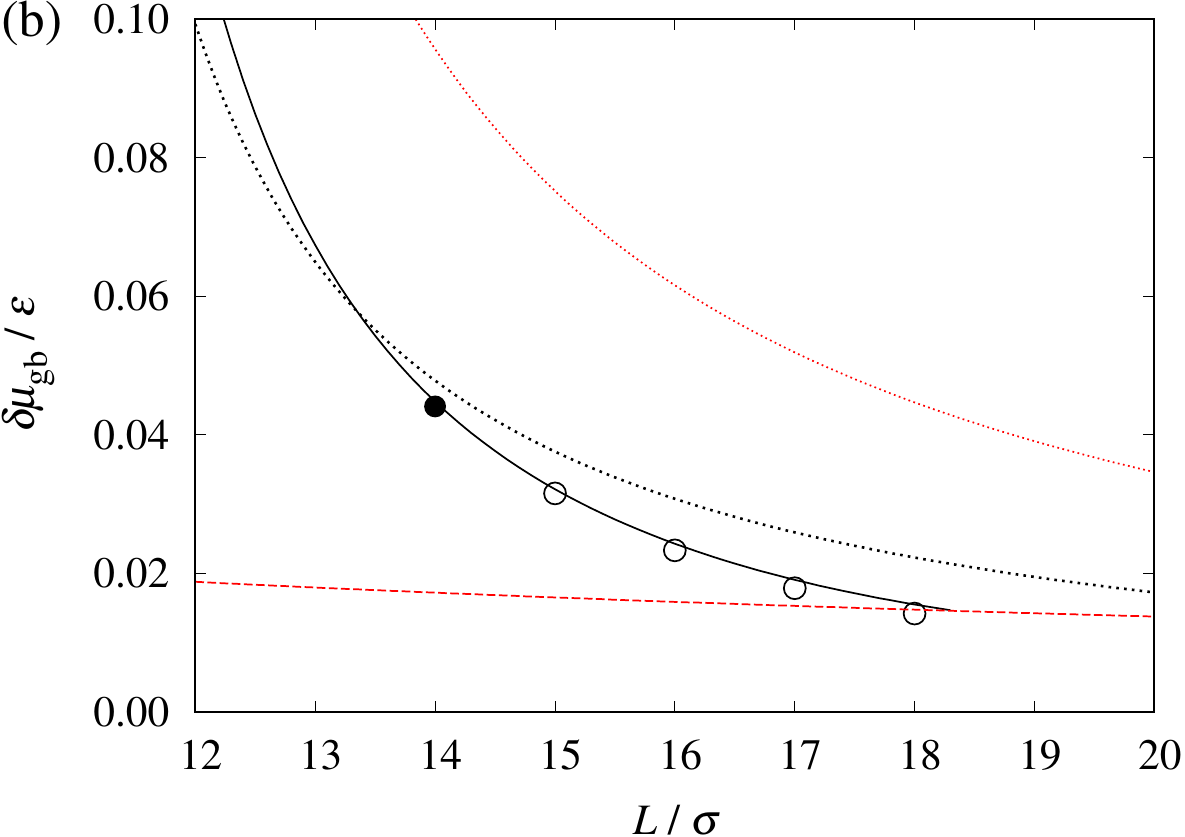}
\caption{Comparison of the location of G-B transition, $\delta\mu_{\rm gb}$, as a function of $L$ obtained from DFT (symbols) and the macroscopic
prediction given by Eq.~(\ref{gas-bridge}) (solid line) for nanopores formed by sinusoidally shaped walls with the amplitude $A=2\,\sigma$ (a) and
$A=5\,\sigma$ (b) and period $P=50\,\sigma$. Also shown are the estimated lower (red dotted line) and upper (red dashed line) spinodals of B phase,
as obtained from Eqs.~(\ref{spin5}) and (\ref{spin6}), respectively. The DFT results include states where the bridges are stable (full circles), as
well as the states where the bridges are metastable (open circles).}\label{gb}
\end{figure}

\begin{figure}[h]
\includegraphics[width=0.45\textwidth]{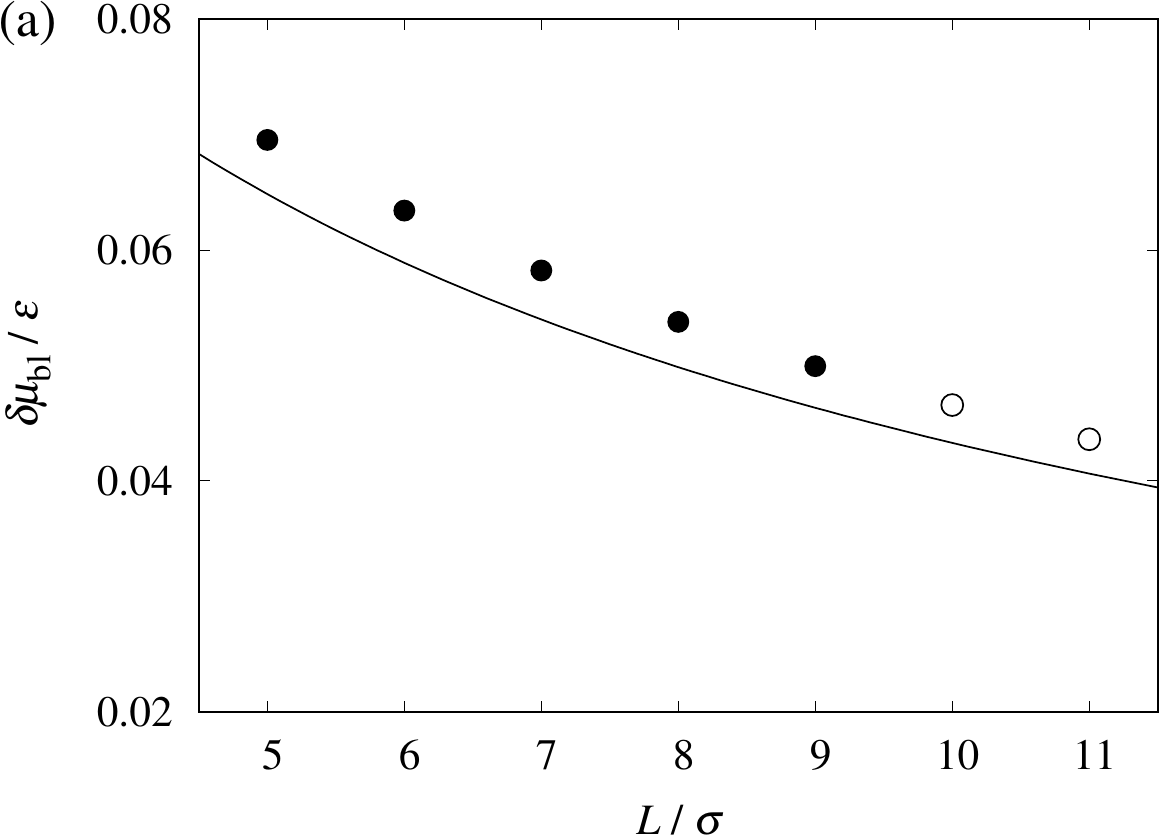}
\includegraphics[width=0.45\textwidth]{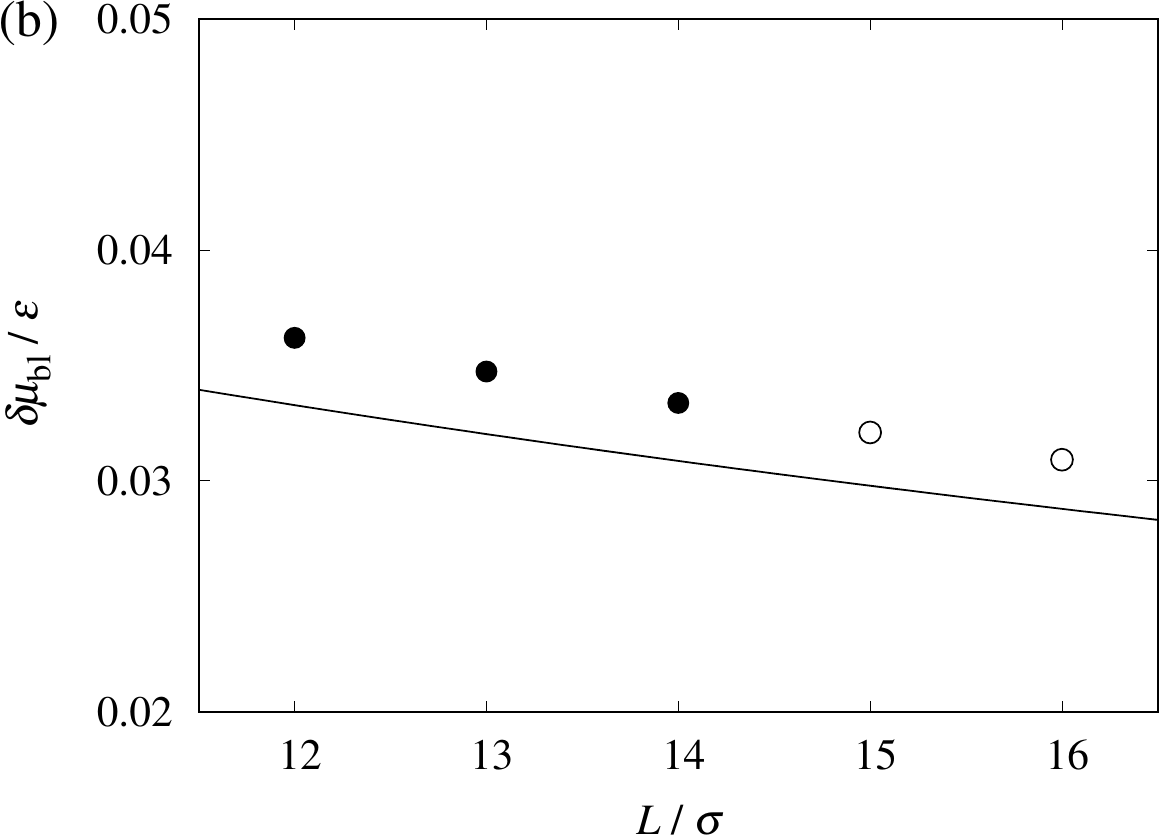}
\caption{Comparison of the location of B-L transition, $\delta\mu_{\rm bl}$, as a function of $L$ obtained from DFT (symbols) and the macroscopic
prediction given by Eq.~(\ref{bridge-liquid}) (line) for nanopores formed by sinusoidally shaped walls with the amplitude $A=2\,\sigma$ (a) and
$A=5\,\sigma$ (b) and the period $P=50\,\sigma$. The DFT results include states where the bridges are stable (full circles), as well as the states
where the bridges are metastable (open circles). }\label{bl}
\end{figure}

Next, instead of varying $a$,  the aspect ratio (and $L$) will be kept constant, such that $A$ and $P$ are varied simultaneously. In
Fig.~\ref{cc_fix_a} we show DFT results for $\mu_{\rm cc}$ as a function of $A$ (and $P$) which are compared with the prediction given by the Kelvin
equation (\ref{sin_kelvin}). Recall that  according to the latter, $\mu_{\rm cc}$ depends on $A$ and $P$ only via their ratio and should thus be
constant. It reveals that although $\mu_{\rm cc}$ is indeed almost invariable for sufficiently large values of $A$ and $P$ and approach the limit,
which is rather close to the Kelvin prediction (with the relative difference about $3\%$), we can also detect a microscopic non-monotonic regime
below $A\approx2\,\sigma$. Here, $\mu_{\rm cc}$ somewhat contra-intuitively drops well below $\mu_{\rm cc}^{\parallel}$ meaning that such a
microscopically small roughness prevents the fluid from condensation. However, this result is completely consistent with the recent microscopic
studies which report that molecular-scale roughness may actually worsen wetting properties of substrates, in a contradiction with the macroscopic
Wenzel law \cite{berim, mal_rough, zhou, svoboda}. This can be explained by a growing relevance of repulsive microscopic forces accompanied by strong
packing effects when the surface roughness is molecularly small \cite{mal_rough}. The decrease of $\delta\mu_{\rm cc}$ upon reducing $A$ (and $P$)
terminates when the amplitude is only a fraction of a molecular diameter ($A\approx0.2\sigma$), where it reaches its minimum; for even finer
structure of the wall the roughness becomes essentially irrelevant and $\mu_{\rm cc}$ approaches its planar limit $\mu_{\rm cc}^{\parallel}$, as
expected.

Finally, we test the Kelvin equation by examining the dependence of $\delta\mu_{\rm cc}$ on $L$. In Fig.~\ref{cc_vary_L} we compare the Kelvin
equation with DFT for nanopores with $A=2\,\sigma$ and $A=5\,\sigma$. In both cases the agreement is very good, especially for large $L$. For the
smallest values of $L$ (but still greater than $L_t$, such that the condensation occurs within one step), $\delta\mu_{\rm cc}$ is slightly
underestimated by the Kelvin equation but the agreement is still very reasonable.

\begin{figure}[h]
\includegraphics[width=0.45\textwidth]{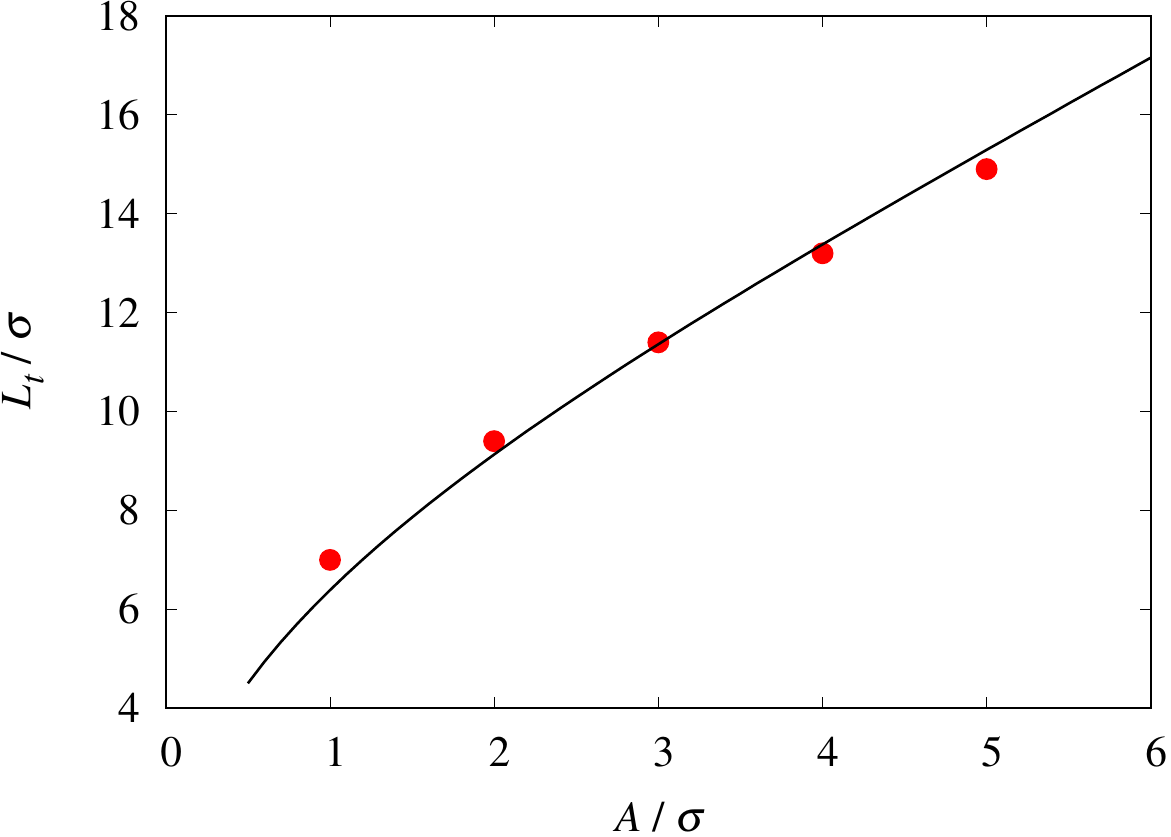}
\caption{Comparison of the threshold mean width $L_t$, allowing for a three-phase coexistence, as a function of the wall amplitude $A$, obtained from
DFT (symbols) and from the macroscopic prediction given by Eq.~(\ref{Lt}) (line). }\label{triple}
\end{figure}

We further consider narrow pores that experience condensation in two steps via formation of liquid bridges. We start with examining the location of
the bridges and test the reliability of Eq.~(\ref{phi}) and its approximative perturbative solution. Fig.~\ref{x0_A2A5} shows a dependence of $x_0$
specifying the location, at which the menisci meet the walls, on $\delta\mu$, as obtained from DFT for nanopores with amplitudes $A=2\,\sigma$ and
$A=5\,\sigma$. The values of $x_0$ corresponding to DFT have been read off from the density profiles in the following way. We approximate the
liquid-gas interface by a part of a circle, $z_c(x)$, of the Laplace radius $R=\gamma/\delta\mu\Delta\rho$. For this, we first determine the point
$(x_m,0)$, where the interface intersects the $x$-axis using the mid-density rule $\rho(x_m,0)=(\rho_g+\rho_l)/2$ (see Fig.~\ref{sketch_complete})
\cite{rule}. This allows us to determine the center of the circle, $x_R=x_m+R$, and the contact point $x_0$ is then obtained using the equal tangent
condition, $z'_w(x_0)=z'_c(x_0)$. The results include the contact points of bridges which correspond both to stable (full symbols) and metastable
(empty symbols) states and are compared with the solutions of the quartic equation (\ref{phi}) and its approximative analytic solution given by
Eq.~(\ref{x0_0}). The comparison shows a very good agreement between DFT and Eq.~(\ref{phi}), which systematically improves with increasing $A$ (as
verified for other models, the results of which are not reported here). This is because the location of bridges is more sensitive to uncertainty in
$R$ for walls with smaller amplitudes. The simple explicit expression (\ref{x0_0}) proves to be a reasonable approximation, except for a near
proximity of saturation; however,  the bridge states are already metastable in this region.

We further test the macroscopic prediction given by Eq.~(\ref{gas-bridge}) for a dependence of $\delta\mu_{\rm gb}$ on $L$. The comparison between
the macroscopic theory and DFT is shown in Fig.~\ref{gb}, again for the amplitudes of $A=2\,\sigma$ and $A=5\,\sigma$. It should be noted that in
both cases the bridging transitions occur over practically identical range of the distance between crests of the opposing walls ($4$--$8\,\sigma$),
although in some cases the transitions lie already in a metastable region. The presence of the lower bound can be interpreted as the minimal width
between the crests allowing for condensation and is comparable with the critical width for the planar slit ($L_c\approx 5\,\sigma$ at this
temperature). On the other hand, the presence of the upper bound  is due to a free-energy cost for the presence of menisci, which destabilizes the
bridges, when $L$ becomes large. The DFT results are compared with the prediction given by Eq.~(\ref{gas-bridge}) (with $x_0$ obtained from
Eq.~(\ref{phi})) and overall the agreement is very good, especially for $A=5\,\sigma$, owing to a very accurate prediction of $x_0$ (cf.
Fig.~\ref{x0_A2A5}). We also plot the estimated lower and upper limits of the bridging states determining the range of stability of bridges for a
given $L$, as obtained from Eqs.~(\ref{spin5}) and (\ref{spin6}). The predicted spinodals indeed demarcate the DFT results for the G-B equilibrium.

\begin{figure}[h]
\includegraphics[width=0.45\textwidth]{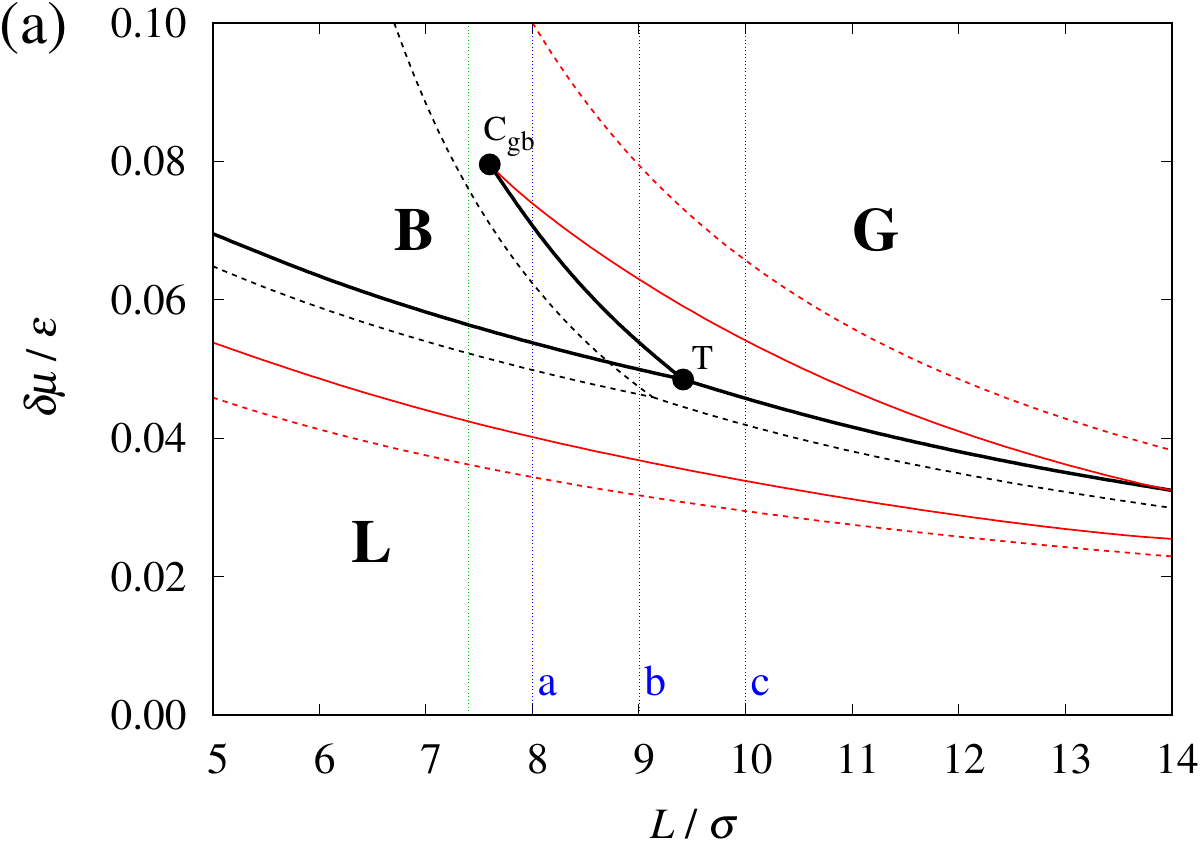}
\includegraphics[width=0.45\textwidth]{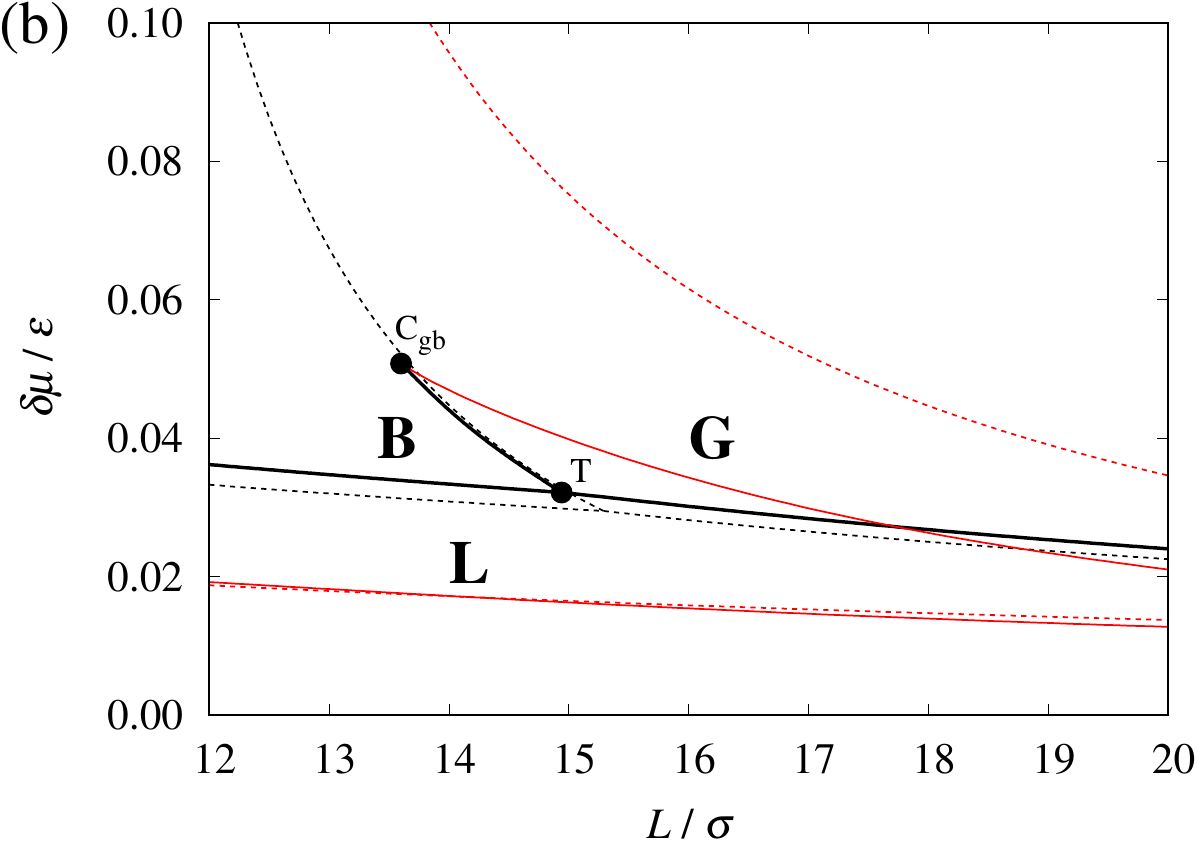}
\caption{Phase diagrams showing the phase behaviour of fluids in nanopores with the walls of amplitudes $A=2\,\sigma$ (a) and $A=5\,\sigma$ (b) and
the period $P=50\,\sigma$, in the $\delta\mu$-$L$ plane. The phase boundaries between G, B and L phases correspond to the DFT results (black solid
line) and the macroscopic theory (black dashed line). Also shown are the spinodals demarcating the limits of stability of B phase, as determined by
DFT (solid red lines) and the macroscopic theory (dashed red lines). All the three phase boundaries meet at the triple point T, for which $L=L_t$
(cf. Fig.~\ref{triple}). The DFT results also include the critical point ${\rm C_{\rm gb}}$, whose presence allows for a continuous formation of
bridges, cf. Fig.~\ref{ads_cont}. The vertical dotted lines depicted in the upper panel correspond to adsorption isotherms shown in Fig.~\ref{ads}
(blue) and in Fig.~\ref{ads_cont} (green).}\label{phase_diag}
\end{figure}

\begin{figure}[h]
\includegraphics[width=0.45\textwidth]{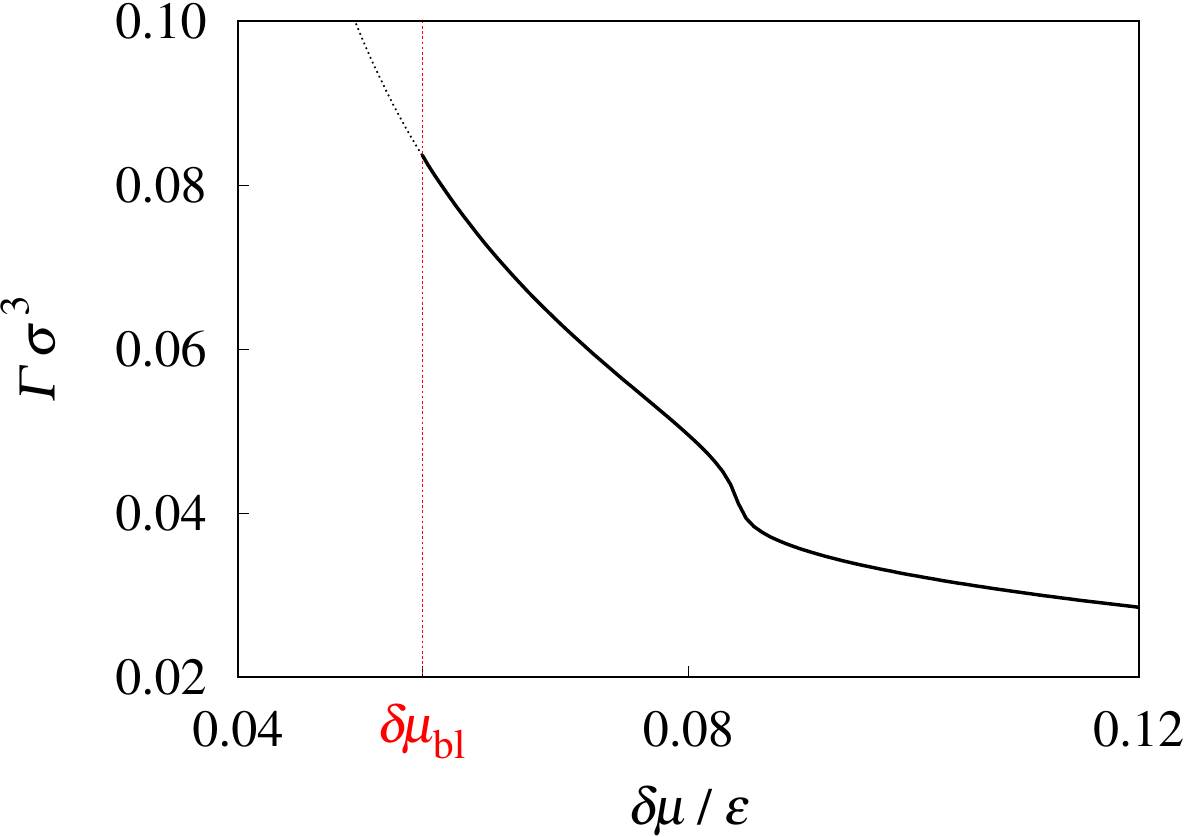}
\caption{Adsorption isotherm corresponding to the nanopore with $A=2\,\sigma$, $P=50\,\sigma$ and $L=7.4\,\sigma$ illustrating continuous formation
of B phase. The thermodynamic path corresponds to the green line in the phase diagram shown in Fig.~\ref{phase_diag}.}\label{ads_cont}
\end{figure}

We now turn our attention to the second step of the condensation process in narrow pores, which corresponds to B-L transition. In Fig.~{\ref{bl} we
compare the dependence of $\delta\mu_{\rm bl}$ on $L$ between DFT results and  the prediction given by Eq.~(\ref{bridge-liquid}). Although still very
reasonable, the agreement, compared to the previous results for G-B transition, is now slightly less satisfactory. This can be attributed to a more
approximative macroscopic description of L phase, which, unlike the low-density G phase, exhibits strongly inhomogeneous structure (cf.
Fig.~\ref{dens_profs_A2_L8}).


In Fig.~\ref{triple} we further show a dependence of $L_t$, separating one-step and two-step condensation regimes, on the wall amplitude $A$. The DFT
results are compared with the macroscopic theory, according to which the dependence of $L_t(A)$ is given implicitly by solving
 \bb
\frac{S_l}{S}=\frac{\ell_w^l-\ell}{\ell_w^l}\,,\;\;\;(L=L_t)\,. \label{Lt}
 \ee
This equation follows by combining any pair of the three phase boundaries conditions, $\delta \mu_{\rm cc}(L)$, $\delta \mu_{\rm gb}(L)$, and $\delta
\mu_{\rm bl}(L)$, as given by Eqs.~(\ref{cc_general}), (\ref{gas-bridge}), and (\ref{bridge-liquid}), respectively. The comparison reveals that the
macroscopic theory is in a close agreement with DFT at least for the considered range of (small) amplitudes.

The phase behaviour in sinusoidal nanopores is summarised in the phase diagrams displayed in Fig.~\ref{phase_diag} for $A=2\,\sigma$ and
$A=5\,\sigma$, where the phase boundaries between G, B and L phases are shown in the $\delta\mu$-$L$ plane. Note that while all the G-L, B-L and B-G
lines terminate at the triple point, only the G-L line is semi-infinite. This is in contrast to the B-L line, which is restricted geometrically by
the condition $L=2A$ and the G-B line which possesses the critical point, allowing for a continuous transition between G and B phases; this is
demonstrated in Fig.~\ref{ads_cont} showing a continuous adsorption corresponding to the green line in  Fig.~\ref{phase_diag}a.  The comparison of
the DFT results with the macroscopic theory reveals an almost perfect agreement for both cases, except for the critical point, which the macroscopic
theory does not capture. Apart from the equilibrium coexistence lines, the borderlines demarcating the stability of the B phase within DFT  are shown
and compared with the lower and upper spinodals according to the geometric arguments (\ref{spin5}) and (\ref{spin6}), respectively. Here, perhaps
somewhat surprisingly, the macroscopic prediction for the upper spinodal is more accurate than for the lower spinodal, especially for the larger
amplitude.

\subsection{$T> T_w$}

\begin{figure}[h]
\includegraphics[width=0.5\textwidth]{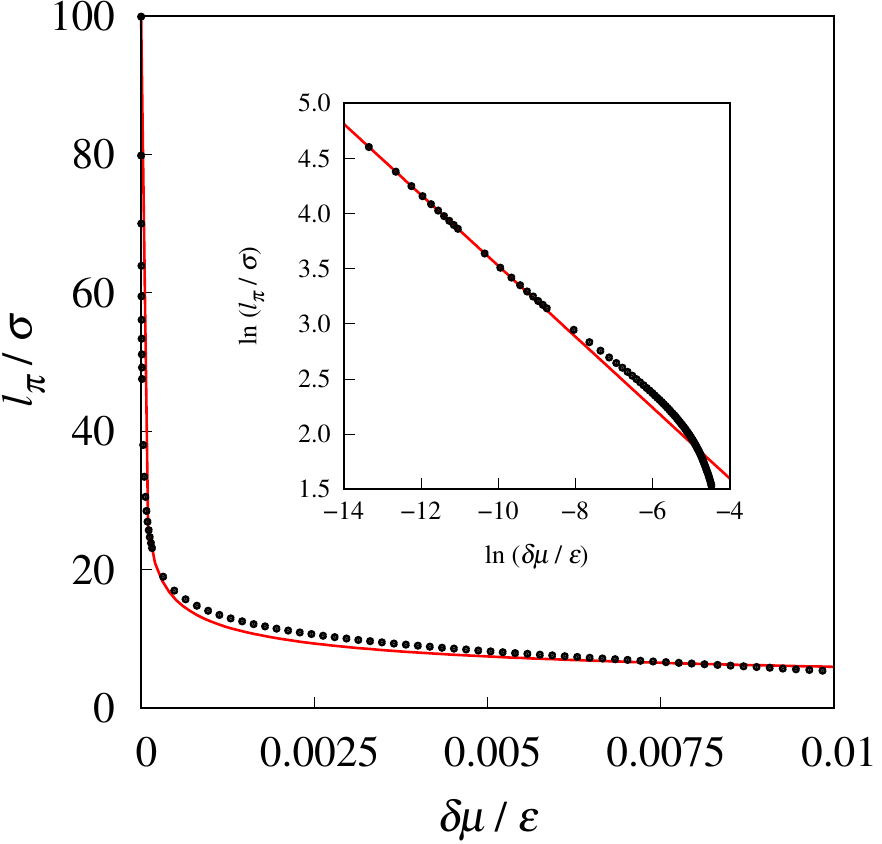}
\caption{DFT results showing the thickness $\ell_\pi$ of the liquid film adsorbed on a planar Lennard-Jones wall as a function of $\delta\mu$. For
small values of $\delta\mu$, the results are consistent with the expected asymptotic power-law, as is verified by the log-log plot shown in the
inset, where the straight line has a slope of $-1/3$. The line in the figure corresponds to the fit of the power-law to the DFT data, which gives
$\ell_\pi=1.363\,\delta\mu^{-1/3}$.}\label{ell_pi}
\end{figure}

\begin{figure}[h]
\includegraphics[width=0.5\textwidth]{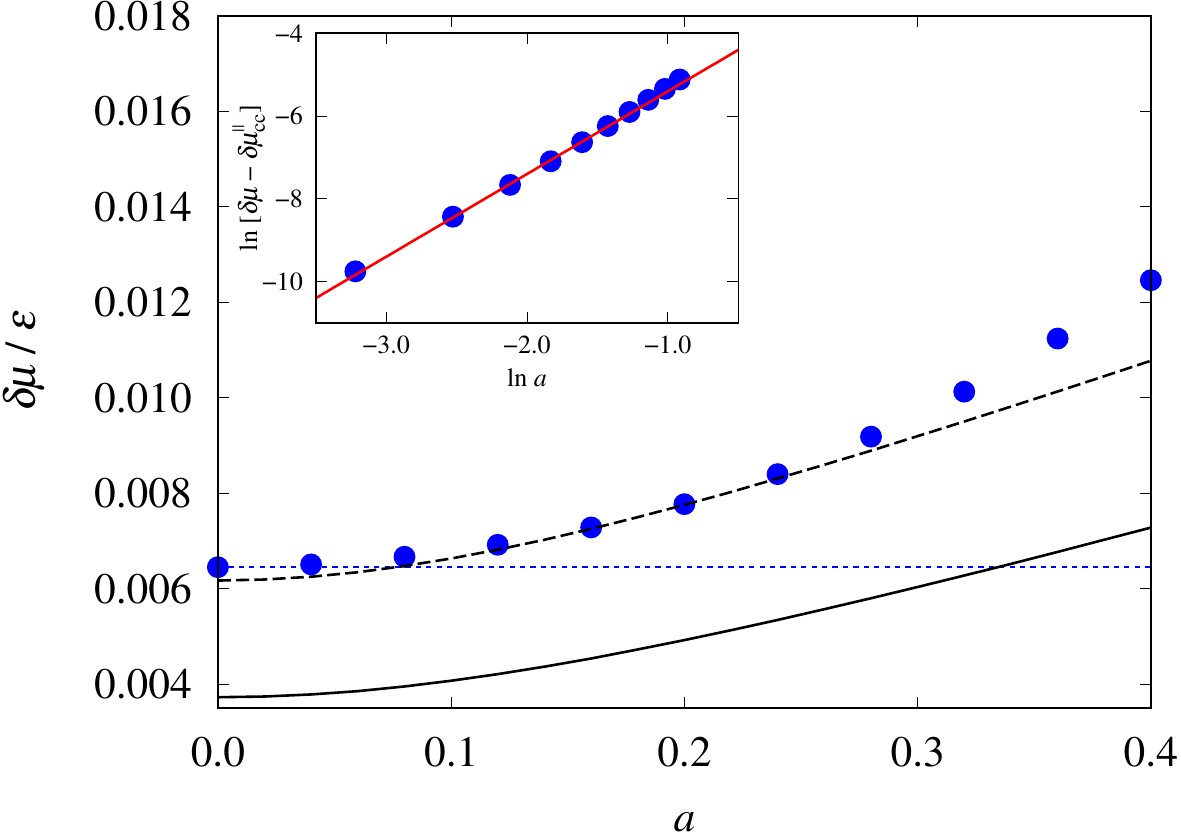}
\caption{Comparison of the dependence of $\delta\mu_{\rm cc}$ on the aspect ratio $a=A/P$ between DFT (symbols), the macroscopic theory,
Eq.~(\ref{sin_kelvin}), (solid line) )and the mesoscopic theory, Eq.~(\ref{sin_kelvin_derj}), (dashed line) for nanopores with $P=50\,\sigma$ and
$L=50\,\sigma$. The dotted line indicates the value of $\delta\mu_{\rm cc}^\parallel$ for capillary condensation in the planar slit obtained from 1D
DFT. The inset shows the log-log plot of the DFT results and the straight line with the slope of $2$ confirms the prediction
\eqref{sin_kelvin2}.}\label{cc_L=50_T=135}
\end{figure}

\begin{figure}[h]
\includegraphics[width=0.5\textwidth]{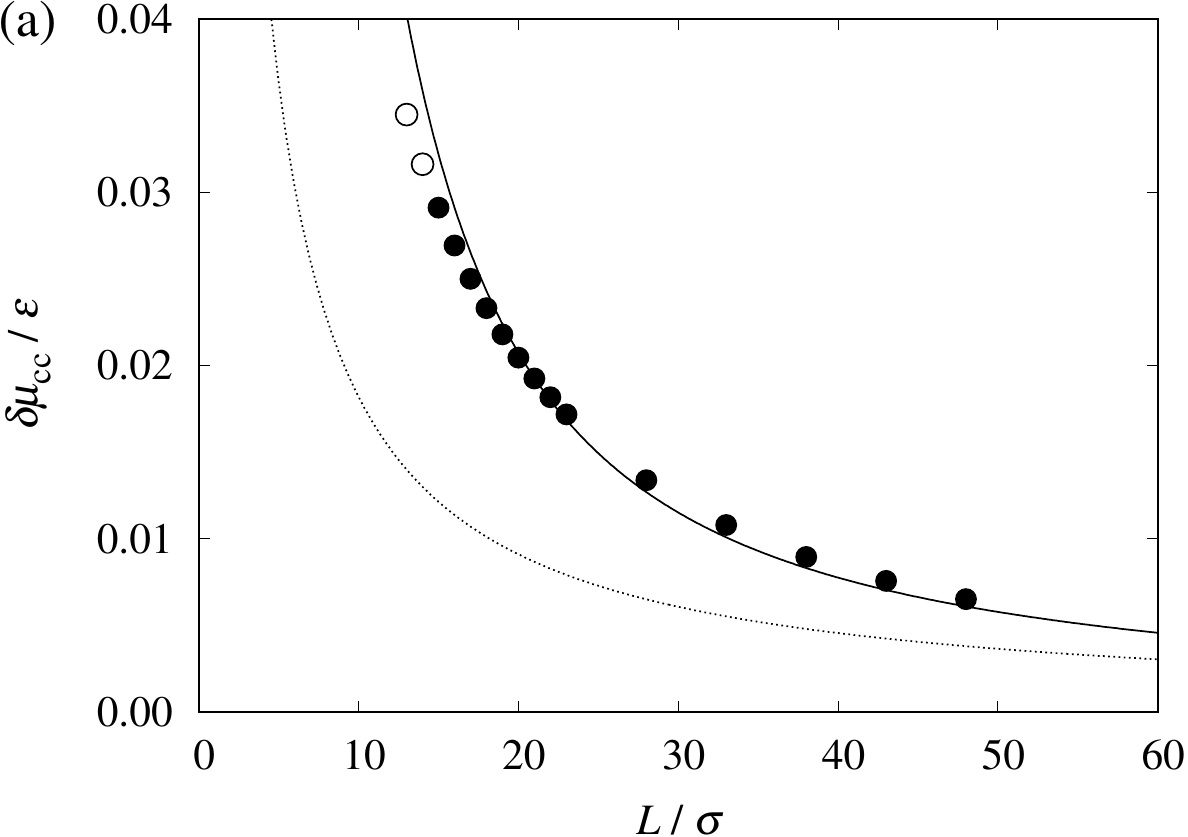}
 \includegraphics[width=0.5\textwidth]{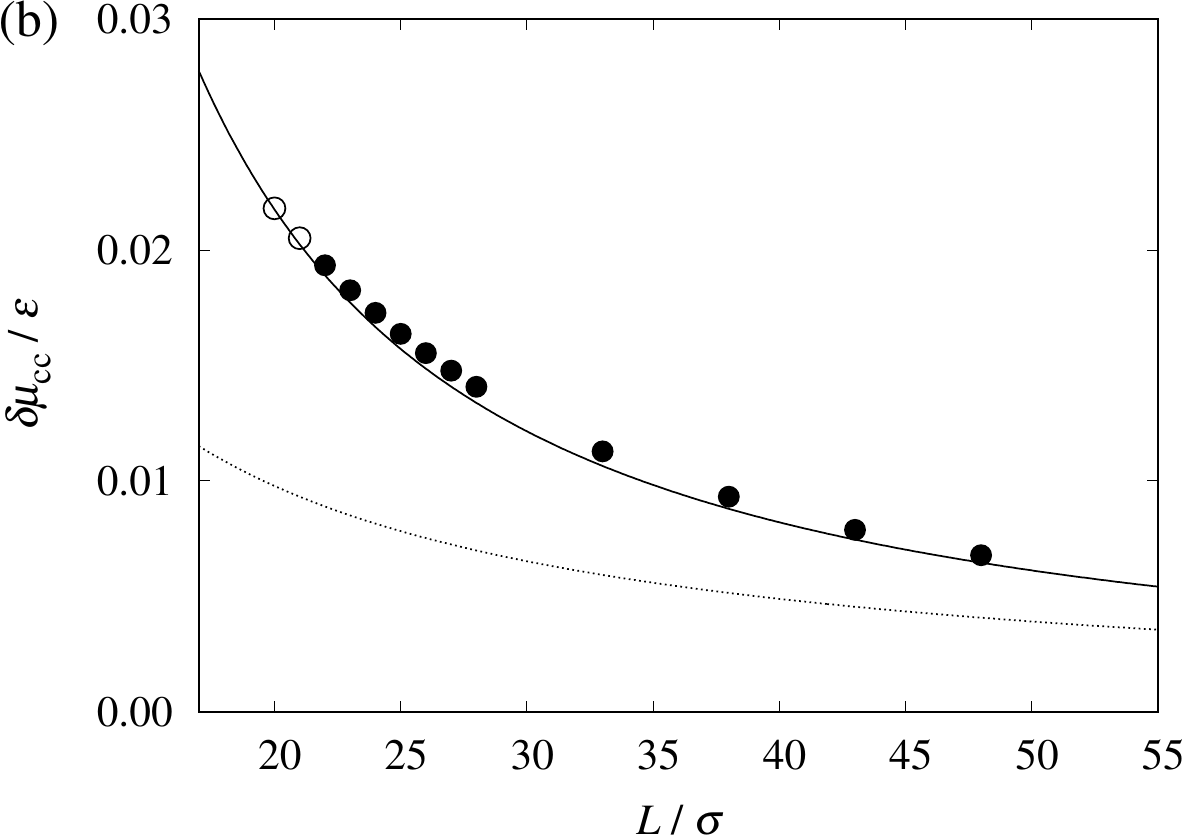}
\caption{Comparison of the dependence of $\delta\mu_{\rm cc}$ on $L$ between DFT results (symbols), the prediction given by the fully macroscopic
Kelvin equation (\ref{sin_kelvin}) (dotted line) and its mesoscopic correction given by  Eq.~(\ref{sin_kelvin_derj}) (solid line).  The nanopores are
formed of sinusoidally shaped walls with the amplitudes of $A=2\,\sigma$ (a) and $A=5\,\sigma$ (b), and period $P=50\,\sigma$. The DFT results
include states which are stable (full circles) and also  metastable (open circles). }\label{cc_dmu-L_T=135}
\end{figure}

Let us now consider a temperature corresponding to $k_BT/\varepsilon=1.35$, which is well above $T_w$, to examine the impact of the wetting layers on
the fluid phase behaviour in sinusoidal nanopores and to test the mesoscopic corrections proposed in section \ref{meso}.  We start by presenting  the
dependence of the film thickness $\ell_\pi$ adsorbed on a planar, 9-3 Lennard-Jones wall, on $\delta\mu$, as obtained from DFT (see
Fig.~\ref{ell_pi}); this is an important pre-requisite for our further mesoscopic analysis requiring an explicit expression for
$\ell_\pi(\delta\mu)$. To this end, we fitted the asymptotic form of $\ell_\pi(\delta\mu)$ to the DFT data obtaining
$\ell_\pi\approx1.363\delta\mu^{-1/3}$. Fig.~\ref{ell_pi} shows that the asymptotic power-law is surprisingly accurate even far from the bulk
coexistence and will thus be used for the further analyzes.

We now turn to wide slits (with $L=50\,\sigma$), for which the condensation is a one-step process from G to L. Fig.~\ref{cc_L=50_T=135} shows the
comparison of DFT results for a dependence of $\delta\mu_{\rm cc}$ on the aspect ratio $a=A/P$, with the predictions obtained from the macroscopic
Kelvin equation, Eq.~(\ref{sin_kelvin}), and its mesoscopic extension given by Eq.~(\ref{sin_kelvin_derj}). While the shape of the graphs
$\delta\mu_{\rm cc}(a)$ given by both theories is very similar, the mesoscopic theory provides a substantial improvement over the macroscopic theory
and yields a near perfect agreement with DFT especially for lower values of $a$. Clearly, the improvement is due to the fact that according to the
mesoscopic theory the nanopores are effectively thinner, which shifts the predicted values of $\delta\mu_{\rm cc}$ upwards (further away from
saturation) compared to the macroscopic treatment. In addition, the horizontal line denoting 1D DFT results for $a=0$ is again completely consistent
with the 2D DFT results, while the inset of the figure confirms the predicted quadratic dependence of $\delta\mu_{\rm cc}$ on $a$ for small values of
the aspect ratio.

Similar conclusion also applies to the results shown in Fig.~\ref{cc_dmu-L_T=135}, where we display a dependence of $\delta\mu_{\rm cc}$ on $L$ for
nanopores with amplitudes $A=2\sigma$ and $A=5\,\sigma$. A comparison between DFT, the macroscopic theory and its mesoscopic correction  is shown for
a large interval of pore widths including those, for which capillary condensation is a two-step process and thus the G-L transition lies in a
metastable region (open circles). In both cases, the mesoscopic correction provides a considerable improvement over the macroscopic theory.

\begin{figure}[h]
\includegraphics[width=0.5\textwidth]{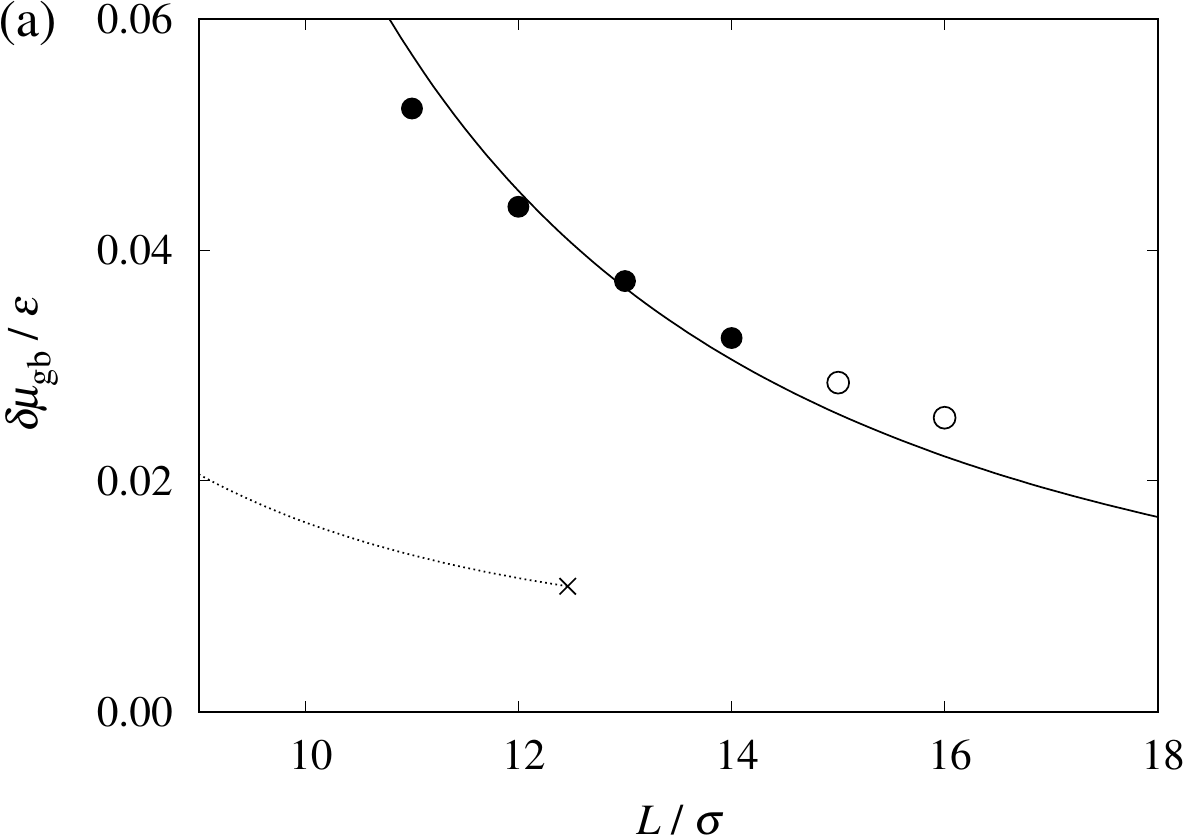}
\includegraphics[width=0.5\textwidth]{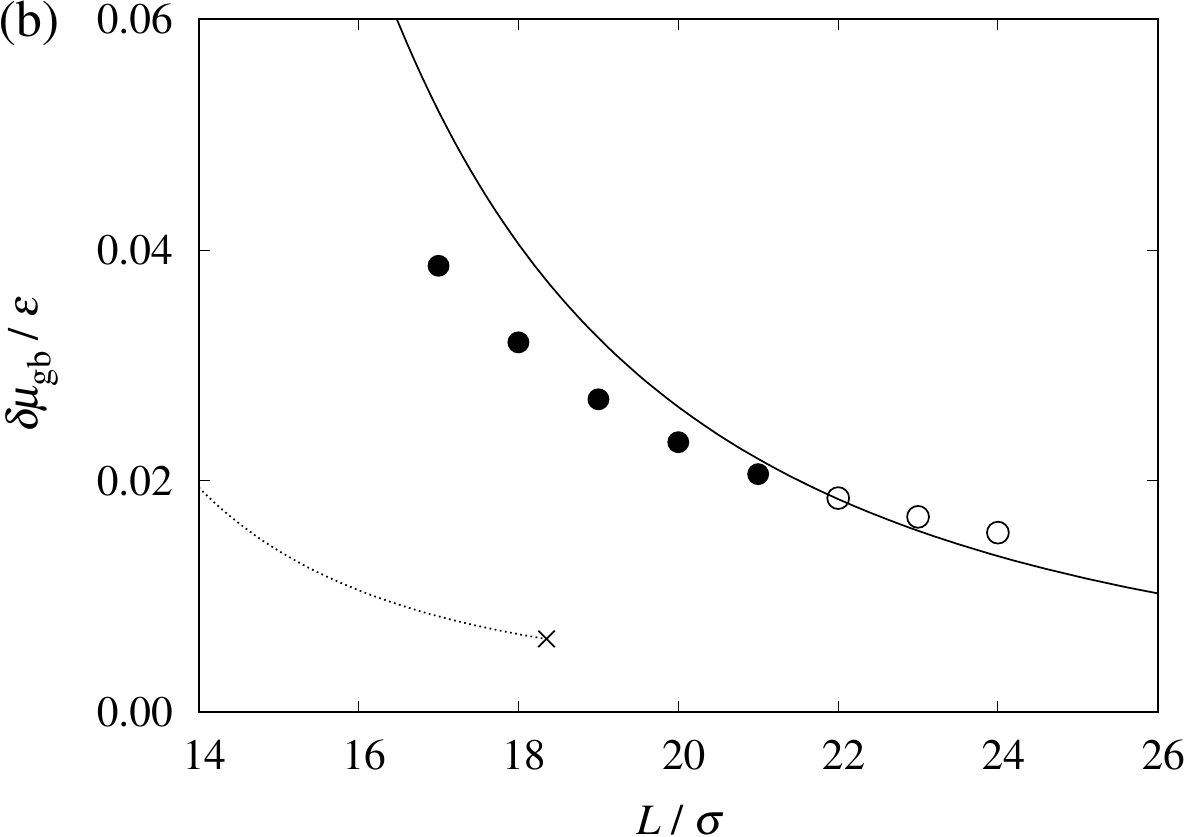}
\caption{Comparison of the location of G-B transition, $\delta\mu_{\rm gb}$, as a function of $L$, obtained from DFT (symbols), the macroscopic
prediction given by Eq.~(\ref{gas-bridge}) (dotted line) and its mesoscopic correction based on the RP construction (full line), for nanopores formed
of sinusoidal walls with the amplitude $A=2\,\sigma$ (a) and $A=5\,\sigma$ (b) and period $P=50\,\sigma$. The macroscopic results terminate at the
(macroscopically predicted) upper limit of B stability (denoted by the cross), when the radius of the bridge menisci becomes $R=R_s^+$. The DFT
results include states which stable (full circles) and also  metastable (open circles). }\label{gb_dmu-L_T=135}
\end{figure}

\begin{figure}[h]
\includegraphics[width=0.5\textwidth]{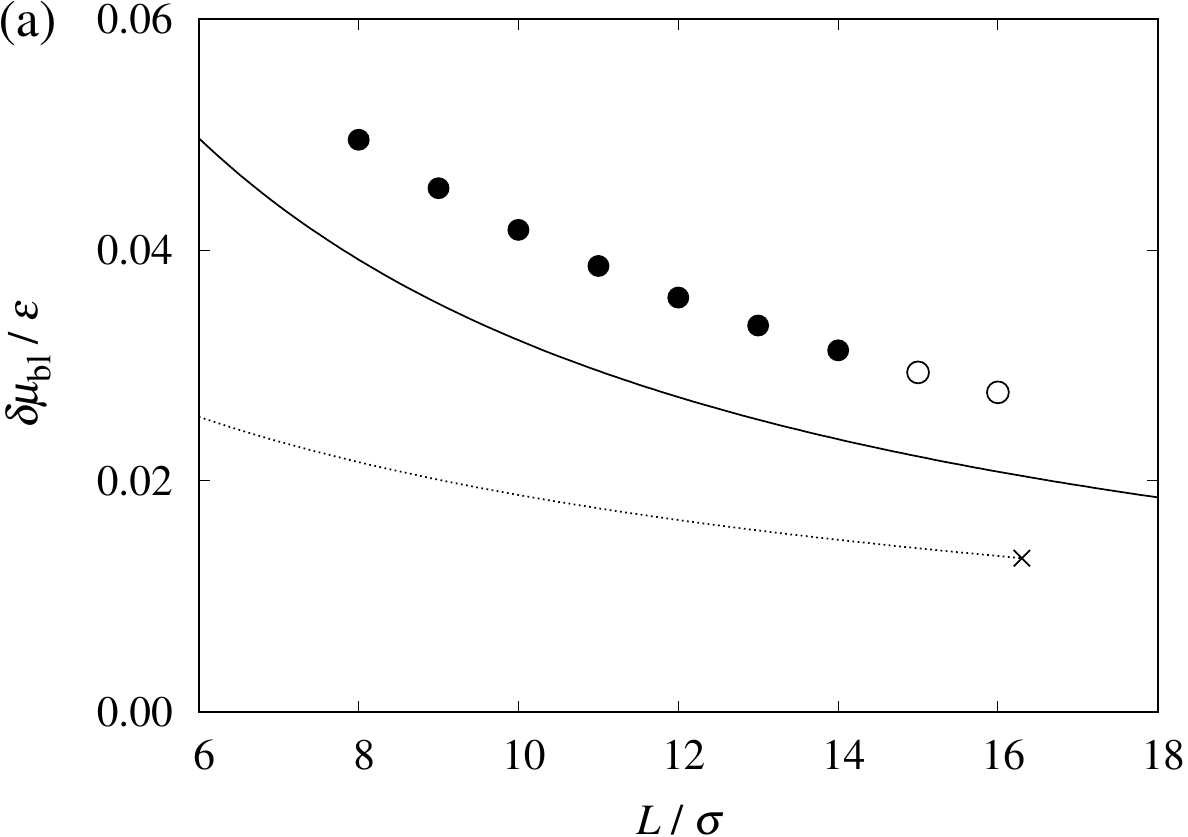}
\includegraphics[width=0.5\textwidth]{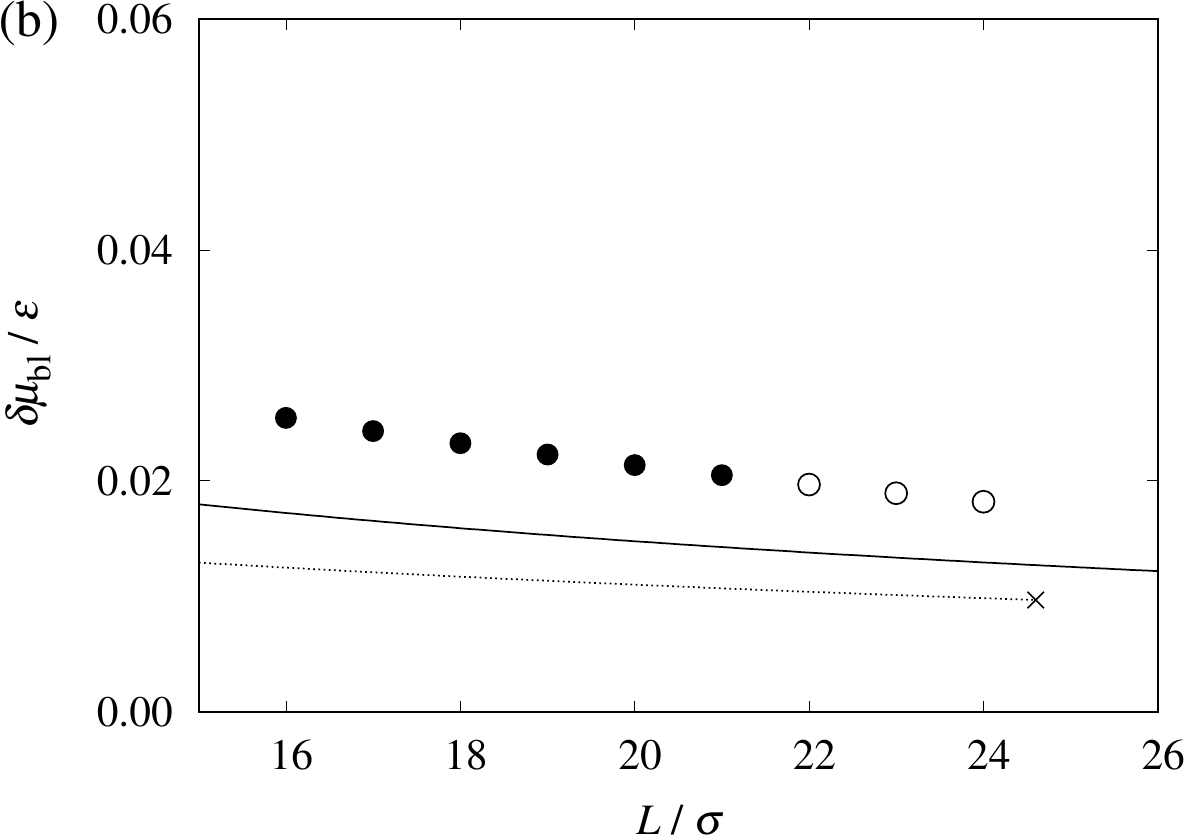}
\caption{Comparison of the location of B-L transition, $\delta\mu_{\rm bl}$, as a function of $L$ obtained from DFT (symbols), the macroscopic
prediction given by Eq.~(\ref{bridge-liquid}) (dotted line) and its mesoscopic correction based on the RP construction (full line), for nanopores
formed of sinusoidal walls with the amplitude $A=2\,\sigma$ (a) and $A=5\,\sigma$ (b) and period $P=50\,\sigma$.  The macroscopic results terminate
at the (macroscopically predicted) lower limit of B stability (denoted by the cross), when the radius of the bridge menisci becomes $R=R_s^-$. The
DFT results include states which stable (full circles) and also  metastable (open circles). }\label{bl_dmu-L_T=135}
\end{figure}

Finally, we test the impact of the mesoscopic correction, now based on the RP construction, for narrow pores, which exhibit capillary condensation in
two steps. The dependence of the location of G-B and B-L transitions on $L$ is shown in Fig.~\ref{gb_dmu-L_T=135} and Fig.~\ref{bl_dmu-L_T=135},
respectively. Again, the mesoscopic correction leads to a remarkable improvement over the macroscopic theory over the entire interval of considered
widths, including those, where G-B and B-L transitions are already metastable w.r.t. to G-L transition. In fact, the improvement is not only
quantitative. It is because that, at this temperature, the macroscopic theory hits the upper spinodal (for the G-B equilibrium) and the lower
spinodal (for the B-L equilibrium) within the range of $L$ where both DFT and the mesoscopic correction allows for the presence of B phase.

\section{Summary and Outlook} \label{summary}

We have studied phase behaviour of fluids confined in nanopores formed by a pair of completely wet walls of smoothly undulated shapes. The varying
local width of such confinements implies that condensation from a low-density phase of capillary gas (G) to a high-density phase of capillary liquid
(L) may be mediated by a sequence of first-order condensation transitions corresponding to a formation of liquid bridges between adjacent parts of
the walls. Our analysis focused on sinusoidally-shaped walls of period $P$ and amplitude $A$, whose mean separation is $L$. The walls are placed such
that one is the reflection symmetry of the other, meaning their local separation varies smoothly between $L-2A$ and $L+2A$. The nature of
condensation in such pores is governed by the mean distance between the walls and can be characterised by the value $L_t$, which is shown to increase
nearly linearly with $A$. For separations $L>L_t$, the condensation is a single-step process from G to L, similar to that in planar slits. However,
for $L<L_t$, the condensation is a two-step process, such that the capillary gas first condenses locally to join the crests of the walls by liquid
bridges forming the bridge phase (B). Upon further increase of the chemical potential (or pressure), the system eventually experiences another
first-order transition corresponding to a global condensation from B to L. It is only for the walls separation $L=L_t$, which allows for a
three-phase G-B-L coexistence.

The phase behaviour of fluids confined by sinusoidal walls has been described in detail using macroscopic, mesoscopic and microscopic models. On a
macroscopic level, we assumed that the confined fluid in G and L phases has a uniform density corresponding to that of a stable bulk gas or a
metastable bulk liquid, at the given temperature and chemical potential. The liquid bridges in B phase are separated from the surrounding gas by
curved menisci, whose shapes were modelled as a part of a circle of the Laplace radius connecting the walls tangentially. Based on this description
we have obtained predictions for the pertinent phase boundaries. Furthermore, we have imposed simple geometric arguments to estimate lower and upper
limits of metastable extensions of B phase.

The comparison with DFT results has shown that the macroscopic description provides a very accurate prediction for the fluid phase behaviour in
sinusoidal pores even for microscopically small values of the geometric parameters, provided the influence of the wetting layers adsorbed at the
walls is insignificant. However, quite generally, their impact cannot be neglected when the pores are formed by completely wet walls of molecularly
small separations. To this end, we have proposed simple mesoscopic corrections of the macroscopic theory, which take into account the presence of the
wetting layers, whose width has been approximated by $\ell_\pi$ corresponding to the film thickness adsorbed on the pertinent planar wall. This
approximation is thus consistent with Derjaguin's correction of the Kelvin equation for the location of capillary condensation in planar slits. For
the transitions involving B phase, we employed the simple geometric construction due to Rasc\'on and Parry, which, too, assumes a coating of the
walls by a liquid film of thickness $\ell_\pi$, which modifies the effective shape and separation of the confining walls. The comparison with DFT
results revealed that the mesoscopic corrections improve the predictions considerably and provide a description of the fluid phase behaviour in
sinusoidally-shaped walls with a remarkable accuracy, at least for the case of low to moderate values of the aspect ratio $a=A/P$.

The reason why we have not considered high values of $a$, is not because the geometric arguments would fail in such cases -- in fact, it was shown
that the predictions for the location of the menisci is more accurate for more wavy walls than for flatter ones -- although the mesoscopic
corrections might be expected to be more approximative as $a$ increases. There is, however, a \emph{qualitative} reason, why the current description
should be modified for such systems. This is related with the phenomenon of the osculation transition \cite{osc} which separates the regimes where
the troughs in G and B phases are filled with a gas (as assumed in this work), from that where the troughs are partially filled with liquid. Allowing
for this phenomenon, and the accompanying interference between the ``vertical'' and the ``horizontal'' menisci, would make the phase behaviour
scenario even much more intricate and we postpone this for future studies.

There are many other possible extensions of this work. For models with high values of $a$, one should also perhaps consider some improvement over the
current mesoscopic corrections that would lead to a geometry- and position-dependent non-uniformity in the width of the wetting layers. A more
comprehensive description of the phase behaviour in sinusoidal nanopores should also take into account the prewetting transition, at which the
thickness of the adsorbed layers has a jump discontinuity. For partially wet walls, the extension of the macroscopic theory would be straightforward
but there is another interfacial phenomenon usually referred to as unbending transition which should be accounted for \cite{unbending}. Natural
modifications of the nanopore model include an examination of the broken reflection symmetry  on the stability of the bridge phase. More intricate
extensions of the current model comprise pair of walls with different wavelengths or walls with additional undulation modes.

\begin{acknowledgments}
\noindent This work was financially supported by the Czech Science Foundation, Project No. 21-27338S.
\end{acknowledgments}


\begin{thebibliography}{99}
\bibitem{four_conv}
We adopoted the following convention for the Fourier transform: $\hat\varphi(\kk)=\int\dd\rr\; \varphi(\rr)\,e^{-2\pi i \kk\cdot\rr}$.
\bibitem{rule}
The location of the interface does not change in any appreciable way if alternative rules, such as the one based on the arithmetic mean of the bulk
packing fractions rather than the densities, is applied.
\end{thebibliography}
\end{document}